\documentclass[pra,twocolumn,groupedaddress,shownopacs,amssymb]{revtex4-2}
\usepackage{graphicx,psfrag,amsmath,amssymb}
\usepackage[usenames, dvipsnames]{color}

\usepackage{hyperref}
\usepackage[normalem]{ulem}
\usepackage{xcolor}
\hypersetup{
    colorlinks=true,
    linkcolor=Blue,
    citecolor=Blue,
    filecolor=Blue,
    urlcolor=Blue,
    pdftitle={SOC-LengthScales paper},
    }
\begin{document}

\title{Pair size versus coherence length and quantum geometry
\\ in the spin-orbit coupled BCS-BEC crossover}

\author{R. N. Kalkan and M. Iskin}

\affiliation{
Department of Physics, Ko\c{c} University, Rumelifeneri Yolu,
34450 Sar\i yer, Istanbul, T\"urkiye
}

\date{\today}

\begin{abstract}

We investigate the interplay between pairing correlations and quantum geometry
in the spin-orbit-coupled BCS-BEC crossover. Working in the helicity basis,
we derive the exact pair-size tensor for the two-body bound states and show
that it separates into an intraband contribution and a quantum-geometric
interband contribution governed by the quantum metric. We extend this
decomposition to two distinct many-body length scales at zero temperature:
the Cooper-pair size from mean-field BCS theory and the coherence length
from Gaussian fluctuation theory. Although the two almost coincide in the
BCS regime, the former measures the internal extent of a pair while the
latter measures the long-wavelength response of the order parameter.
Evaluating both for three-dimensional Rashba, three-dimensional Weyl, and
two-dimensional Rashba spin-orbit couplings, we find that the pair size
decreases monotonically with spin-orbit coupling, whereas the coherence
length develops a pronounced minimum, associated with the collapse of
the noninteracting Fermi surface onto the ring or sphere of
helicity-band minima, before growing again as the composite bosons become weakly
interacting. The quantum-geometric contributions reach up to about 30\%
of the squared pair size and squared coherence length, tracking the same
changes in the helicity Fermi-surface topology.

\end{abstract}

\maketitle

\section{Introduction}
\label{sec:intro}

The BCS-BEC crossover provides a unified framework for describing fermionic
superfluidity across the full range of pairing strengths, from weakly-bound
Cooper pairs in the BCS regime to tightly-bound diatomic molecules undergoing
BEC~\cite{inguscio07, zwerger11, zhai21}.
Ultracold Fermi gases near a Feshbach
resonance have proven to be an exceptionally clean and tunable platform for
realizing this crossover, allowing the interaction strength to be varied
continuously via an external magnetic
field~\cite{ketterle08, strinati18}. Among the many
observables used to characterize the crossover, two length scales play a
central role: the pair size, which measures the internal spatial extent of a
fermion pair, and the coherence length, which measures the distance over
which the superfluid order parameter remains phase coherent. While these two
quantities are often used interchangeably in the BCS regime,
where they coincide up to numerical prefactors, it has long been appreciated
that they encode distinct physical information away from it and,
in general, need not be
equal~\cite{sademelo93, engelbrecht97, pistolesi94, pistolesi96}.

A further ingredient has become available with the experimental realization
of synthetic spin-orbit coupling (SOC) in ultracold atomic
gases~\cite{galitski13, zhai15, zhang19}.
Using Raman-dressing schemes, SOC of the Rashba and Rashba-Dresselhaus type
has been engineered in bosonic and fermionic quantum gases, and subsequent
experiments have extended these techniques to 2D and 3D SOC geometries,
including proposals and implementations
approaching isotropic Weyl-type coupling~\cite{cheuk12, meng16, huang16}.
These developments have opened the door to studying the interplay of pairing
and SOC directly in the laboratory, with measurements of the modified
Fermi-surface (FS) topology, the closing and reopening of pairing gaps,
and signatures of the crossover from BCS pairing to a BEC of
tightly-bound ``rashbons''~\cite{vyasanakere12}, whose coherence
properties we compare against effective Gross-Pitaevskii theory below.
On the theoretical side, SOC is known to reshape the single-particle
band structure into helicity bands with momentum-dependent spinors,
fundamentally altering the pairing problem relative to the SOC-free case.

A key theoretical consequence of this momentum-dependent spinor structure is
the emergence of quantum geometry. The helicity eigenstates vary with
momentum in a way that is captured by the quantum-geometric tensor, whose
real part defines the quantum metric and whose imaginary part is related
to the Berry curvature~\cite{Provost80, resta11, torma23}.
The quantum metric has recently been probed
experimentally in spin-momentum-locked systems and 3D
topological insulators, underscoring that this geometric structure
is not merely a theoretical construct but a directly measurable
property of spin-orbit-coupled systems~\cite{sala25,sala26}.
In spin-orbit-coupled Fermi superfluids, this
quantum metric has already been shown to leave a direct and quantifiable
imprint on several many-body properties: it contributes an anomalous
geometric piece to the superfluid density, it enters the effective-mass
tensor governing the dispersion of Cooper pairs, and it modifies the
velocity of the Goldstone mode~\cite{iskin18a, iskin18b, iskin20a}.
These results establish that virtual interband transitions between
helicity bands are not a small correction but a structurally distinct
geometric contribution that adds to the conventional intraband
physics inherited from the band dispersion alone. Although only the
total of the two contributions is measured, separating them by their
physical origin, as is routinely done for the conventional and geometric
parts of the superfluid weight~\cite{torma22, peotta23}, is what
identifies which features of an observable follow from the band
dispersion and which from the geometry of the Bloch or helicity states.
The pair size and the coherence length are natural next quantities
to examine from this perspective, since both are built from the same
underlying pairing correlations and might therefore be expected to
inherit a comparable quantum-geometric contribution; whether they do so
in the same way, however, has not been established, and it is this
question that we address in this paper.

This question is sharpened by a separate ongoing controversy in the
flat-band superconductivity literature, where the relation between the
pair size and the coherence length has become a point of active
disagreement~\cite{hu23, thumin24, iskin24c, iskin25, elden26}.
In flat-band systems, the absence of single-particle dispersion means
that superfluidity is controlled entirely by quantum
geometry~\cite{torma22, peotta23, yu25, liu25, gao25}, and different
works have reached seemingly conflicting conclusions about whether the
coherence length can be shorter than, equal to, or unrelated to the size
of a single Cooper pair, and about what such a comparison even means when
the kinetic energy vanishes identically~\cite{li25, virtanen25, lee26,
xiao25, oh25, chen26, ozkurt26}. Precisely because quantum geometry
dominates completely there, with no intraband contribution left to
compare against, the distinct roles of the pair size and the coherence
length are difficult to isolate in flat bands. The spin-orbit-coupled
Fermi gas offers a complementary and more tractable setting for the same
question: quantum geometry enters through a tunable interband
contribution that coexists with, rather than replaces, a finite
intraband dispersion, and the interaction strength and SOC can be dialed
continuously from the two-body limit through the full crossover. A clean
understanding of how and why the pair size and coherence length differ
in this simpler continuum setting may therefore help clarify which
features of the flat-band controversy are intrinsic to flat bands, and
which reflect a more general distinction between two length scales that
have not always been kept apart in the literature.

In this paper, we formulate the spin-orbit-coupled BCS-BEC crossover
problem in the helicity basis and derive the exact pair-size tensor for
the two-body bound states, showing that it decomposes into an intraband
contribution set by the helicity-band dispersion and a quantum-geometric
interband contribution governed entirely by the quantum metric. We extend this
decomposition to the many-body Cooper-pair size within mean-field BCS theory
and to the zero-temperature ($T = 0$) coherence length obtained from the
Gaussian fluctuation theory, and we evaluate both quantities
numerically across the interaction-SOC parameter space for
3D Rashba, 3D Weyl, and 2D Rashba SOC models. We find that the pair size
and coherence length, despite almost coinciding in the BCS regime,
exhibit qualitatively different dependences on SOC strength, and we
quantify the fraction of each squared length that originates from
quantum geometry. This geometric contribution is not merely a
correction: it can change the qualitative dependence of the coherence
length on the SOC strength and, in two dimensions, it changes sign,
shortening the coherence length below its band-structure value.
We also clarify how these two length scales relate to the pairing-related
lengths previously introduced for spin-orbit-coupled Fermi gases: the
quantity called the coherence length in Ref.~\cite{yu2014coherence}
coincides at the mean-field level with the Cooper-pair size defined
here (denoted $\xi_\mathrm{Cp}$ below), whereas the fluctuation-based
coherence length (denoted $\xi_0$ below) has no counterpart in earlier
studies of these systems.

The remainder of this paper is organized as follows.
Section~\ref{sec:bcsbec} develops the helicity-basis formalism and
derives the intraband-interband decomposition of the two-body pair size,
the Cooper-pair size, and the coherence length.
Section~\ref{sec:numerical_results} presents the numerical results for
the three SOC models and compares them across geometries in
Sec.~\ref{sec:comparison}, Sec.~\ref{sec:literature} clarifies the physical
distinctions between the length scales defined here and those used in
the literature, and Sec.~\ref{sec:conc} concludes with a brief outlook.

\section{BCS-BEC crossover theory}
\label{sec:bcsbec}

We start by formulating the spin-orbit-coupled Hamiltonian in the helicity
basis, where the quantum metric naturally emerges and provides the
geometric framework underlying the results that follow.

\subsection{Continuum Hamiltonian}
\label{sec:single_particle}

Having in mind a uniform spin-$1/2$ Fermi gas in $d = \{2,3\}$ spatial
dimensions, we expand the fermionic field operators in plane-wave
modes as
$
\psi_\sigma(\mathbf{r})=\frac{1}{\sqrt V}
\sum_{\mathbf{k}}e^{i\mathbf{k}\cdot\mathbf{r}} c_{\sigma \mathbf{k}},
$
where $\sigma = \{\uparrow,\downarrow\}$ labels the spins,
$V$ is the area/volume of the system, $\mathbf k$ is the
momentum in units of $\hbar = 1$, and $c_{\sigma \mathbf{k}}$ annihilates
a fermion with spin $\sigma$ and momentum $\mathbf{k}$. Defining the spinor
$
\psi_{\mathbf{k}}=(c_{\uparrow \mathbf{k}}, \,
c_{\downarrow \mathbf{k}})^\mathrm{T},
$
the single-particle Hamiltonian can be written as
\begin{equation}
\mathcal{H}_0=\sum_{\mathbf{k}}\psi_{\mathbf{k}}^{\dagger}\left[\xi_{\mathbf{k}}\mathbb{I}+\mathbf{d}_{\mathbf{k}}\cdot\boldsymbol{\sigma}\right]\psi_{\mathbf{k}},
\label{eq:single_particle_hamiltonian_momentum}
\end{equation}
where
$
\xi_{\mathbf{k}} = \varepsilon_\mathbf{k} - \mu
$
with $\varepsilon_\mathbf{k} = k^2/(2m)$ and $k = |\mathbf{k}|$ is the
free-particle dispersion measured relative to the chemical potential
$\mu$, $\mathbb I$ is the $2\times2$ identity matrix in spin space,
$
\boldsymbol{\sigma}=(\sigma_x,\sigma_y,\sigma_z)
$
is the vector of Pauli matrices, and
$
\mathbf{d}_{\mathbf{k}} = (d_{\mathbf{k}}^x,d_{\mathbf{k}}^y,d_{\mathbf{k}}^z)
$
is the SOC field. In this paper we focus on three SOC models; we do not
consider the 1D case, since a purely 1D SOC field can always
be removed by a local spin rotation and therefore has no observable effect.
For the 3D Rashba model, $\mathbf k=(k_x,k_y,k_z)$ and the SOC field acts
only in the $xy$ plane,
$
\mathbf d_{\mathbf k}=\alpha(k_y,-k_x,0).
$
For the 3D Weyl model, we consider the isotropic SOC field
$
\mathbf d_{\mathbf k}=\alpha(k_x,k_y,k_z).
$
Finally, for the 2D Rashba model, where $\mathbf k=(k_x,k_y)$, we consider the SOC field
$
\mathbf d_{\mathbf k}=\alpha(k_y,-k_x,0).
$
Here, $\alpha$ denotes the SOC strength and has dimensions of velocity.

Next we transform from the spin basis to the helicity basis.
The helicity states $|u_{s\mathbf{k}}\rangle$ are labeled by $s=\pm1$,
where $s=+1$ corresponds to spin polarization locally aligned with the SOC
field $\mathbf d_{\mathbf k}$ and $s=-1$ corresponds to spin polarization
locally anti-aligned with it. These states diagonalize the SOC term in
the Hamiltonian according to
\begin{align}
\mathbf{d}_{\mathbf{k}}\cdot\boldsymbol{\sigma}|u_{s\mathbf{k}}\rangle
= s\, d_\mathbf{k}|u_{s\mathbf{k}}\rangle,
\end{align}
where $d_\mathbf{k} = |\mathbf{d}_\mathbf{k}|$. The corresponding helicity
dispersions are $\xi_{s\mathbf{k}}=\xi_{\mathbf{k}}+s d_{\mathbf{k}}$, so
that the SOC splits the original band into two helicity branches separated
in energy by $2d_{\mathbf{k}}$. Writing the spin-basis components of the
helicity states as $u_{s\mathbf{k}}^\sigma = \langle\sigma|u_{s\mathbf{k}}\rangle$,
the spin and helicity annihilation operators are related by
$
c_{\sigma \mathbf{k}}=\sum_{s}u_{s\mathbf{k}}^\sigma c_{s\mathbf{k}},
$
where $c_{s\mathbf{k}}$ annihilates a fermion of helicity band $s$ at momentum
$\mathbf{k}$. The single-particle Hamiltonian thus takes the diagonal form
\begin{equation}
\mathcal{H}_0=\sum_{s\mathbf{k}} (\varepsilon_{s\mathbf{k}} - \mu)
c_{s\mathbf{k}}^{\dagger} c_{s\mathbf{k}}, \quad
\varepsilon_{s\mathbf{k}}=\varepsilon_{\mathbf{k}}+s d_{\mathbf{k}},
\label{eq:single_particle_hamiltonian_helicity}
\end{equation}
in the helicity basis.

For the 3D Weyl and 2D Rashba models, the $-$-helicity dispersion
takes the form
$
\xi_{-,\mathbf k}=(k-m\alpha)^2/(2m)-\mu-m\alpha^2/2.
$
The degenerate minima at $k=m\alpha$ form a sphere in the 3D Weyl model
and a ring in the 2D Rashba model. For the 3D Rashba model,
$
\xi_{-,\mathbf k}=[(k_\perp-m\alpha)^2+k_z^2]/(2m)-\mu-m\alpha^2/2,
$
where $k_\perp=\sqrt{k_x^2+k_y^2}$. In this case, the degenerate minima
at  $k_\perp=m\alpha$ and $k_z=0$ form a ring in the $k_xk_y$ plane.
Thus, for the noninteracting system, the FS structure
changes at two characteristic values of the chemical potential, $\mu=0$ and
$\mu=-m\alpha^2/2$ (see Table~\ref{tab:fermi_surface})~\cite{cappelluti07}.
At $\mu=0$, the $+$-helicity FS shrinks to the point $\mathbf k= \mathbf 0$
and disappears for $\mu<0$.
In the interval $-m\alpha^2/2<\mu<0$, only the $-$-helicity band crosses
the chemical potential. The corresponding FS consists of
two spherical surfaces for the 3D Weyl model,
two circular contours for the 2D Rashba model, and
one toroidal surface for the 3D Rashba model. At $\mu=-m\alpha^2/2$,
the chemical potential reaches the bottom of the $-$-helicity band.
The two spherical surfaces merge into the sphere of minima in the 3D
Weyl model, the two circular contours merge into the ring of minima in
the 2D Rashba model, and the toroidal surface contracts into the ring of
minima in the 3D Rashba model. For $\mu < -m\alpha^2/2$, no noninteracting
FS remains. These two characteristic points therefore mark distinct
changes in the low-energy helicity-band structure and are relevant to
the features observed in the pair-size and coherence-length calculations
of Sec.~\ref{sec:numerical_results}. As we discuss in
Sec.~\ref{sec:twobody}, the diverging density of states on this ring or
sphere of minima also has a direct consequence for two-body binding in
the strong-SOC limit.

\begin{table}[htb]
\begin{tabular}{c|c|c}
\hline
SOC & $\mu=0$ & $\mu=-m\alpha^2/2$\\
\hline
3D Weyl   & $+$ FS disappears & $-$ FSs $\to$ sphere\\
3D Rashba & $+$ FS disappears & $-$ FSs $\to$ ring\\
2D Rashba & $+$ FS disappears & $-$ FSs $\to$ ring\\
\hline
\end{tabular}
\caption{Characteristic Fermi-surface (FS) changes in the
noninteracting SOC Fermi gases.}
\label{tab:fermi_surface}
\end{table}

The FS changes described above concern only the energetics of the
helicity bands; the corresponding states $|u_{s\mathbf k}\rangle$ vary
continuously with $\mathbf k$ as well. To characterize this momentum-space
variation, we introduce the quantum-geometric
tensor~\cite{Provost80, resta11, torma23},
$
\mathcal Q^{s\mathbf k}_{ij} = \langle\partial_i u_{s\mathbf k}|
\left(\mathbb I-P_{s\mathbf k}\right)
|\partial_j u_{s\mathbf k}\rangle,
$
where
$
P_{s\mathbf k}=|u_{s\mathbf k}\rangle\langle u_{s\mathbf k}|
$
is the projector onto helicity band $s$ and $\partial_i=\partial/\partial k_i$.
Its real symmetric part defines the quantum metric, while its imaginary
antisymmetric part is related to the Berry curvature. With the normalization
used here, the quantum metric of helicity band $s$ is
$
g^{s\mathbf k}_{ij}=2\mathrm{Re}\mathcal Q^{s\mathbf k}_{ij}.
$
Using the completeness of the helicity states, the quantum metric of
band $s$ can be decomposed into contributions from the other helicity bands
as
$
g^{s\mathbf k}_{ij}=\sum_{s'\neq s}g^{ss'\mathbf k}_{ij},
$
where
$
g_{ij}^{ss'\mathbf k}=2\mathrm{Re}\left[
\langle\partial_i u_{s\mathbf k}|u_{s'\mathbf k}\rangle
\langle u_{s'\mathbf k}|\partial_j u_{s\mathbf k}\rangle
\right]
$
is the band-resolved quantum-metric tensor describing the contribution of
band $s'$ to the quantum metric of band $s$. For the two-band helicity
Hamiltonian considered here, the only other band is $s'=-s$, and therefore
$
g^{s\mathbf k}_{ij}=g^{s,-s,\mathbf k}_{ij}.
$
Defining the unit vector
$
\hat{\mathbf d}_{\mathbf k}=\mathbf d_{\mathbf k}/d_{\mathbf k},
$
the quantum metric reduces to
\begin{equation}
g^{\mathbf k}_{ij}=\frac{1}{2} \partial_i\hat{\mathbf d}_{\mathbf k}\cdot\partial_j\hat{\mathbf d}_{\mathbf k},
\label{eq:quantum_metric_dvector}
\end{equation}
which is independent of the helicity index $s$.
Equation~\eqref{eq:quantum_metric_dvector} shows explicitly that the quantum
metric is governed entirely by the angular variation of the SOC field
$\hat{\mathbf d}_{\mathbf k}$.

As noted above, the $-$-helicity states along the sphere or ring of minima
are energetically degenerate, while the corresponding helicity states remain
momentum-dependent because the orientation $\hat{\mathbf d}_{\mathbf k}$
varies along the manifold. Consequently, the quantum metric remains nonzero
even though the energy is constant along the manifold of minima. This
behavior is analogous to that of exactly flat Bloch bands, where the quantum
geometry can remain nonzero despite the vanishing band dispersion.
In an exactly flat band, however, the energy is independent of momentum
throughout the entire Brillouin zone, whereas the SOC bands exhibit local
flatness only along their lower-dimensional manifolds of minima and remain
dispersive away from them.

Having specified the noninteracting SOC Hamiltonian, we now introduce an
attractive two-body interaction between opposite-spin fermions, giving the
full many-body Hamiltonian
$
\mathcal H = \mathcal H_0 + \mathcal H_{\mathrm{int}}.
$
In momentum space, this interaction becomes
\begin{equation}
\mathcal{H}_{\mathrm{int}}=-\frac{U}{V}\sum_{\mathbf{k} \mathbf{k}' \mathbf{q}}
c^{\dagger}_{\uparrow,\mathbf{k}+\mathbf{q}/2}
c^{\dagger}_{\downarrow,-\mathbf{k}+\mathbf{q}/2}
c_{\downarrow,-\mathbf{k}'+\mathbf{q}/2}
c_{\uparrow,\mathbf{k}'+\mathbf{q}/2},
\label{eq:interaction_hamiltonian_spin_momentum}
\end{equation}
where $U > 0$ is the strength of the contact (zero-range) interaction.
The contact interaction requires ultraviolet regularization.
In 3D, the bare interaction strength $U$ is related to the physical
$s$-wave scattering length $a_s$ through
$
V/U=-mV/(4\pi a_s) + \sum_{\mathbf k}1/(2\varepsilon_{\mathbf k}).
$
In 2D, it is instead related to the two-body binding energy $E_b>0$
in vacuum in the absence of SOC through
$
V/U=\sum_{\mathbf k}1/(2\varepsilon_{\mathbf k} + E_b).
$
Since the contact interaction couples only to the spin-singlet channel,
we introduce the normalized spin-singlet pair creation operator
$
S_{\mathbf{k}}^{\mathbf q\dagger}=\frac{1}{\sqrt{2}}\big(
c_{\uparrow,\mathbf{k}+\mathbf{q}/2}^{\dagger}
c_{\downarrow,-\mathbf{k}+\mathbf{q}/2}^{\dagger}
-
c_{\downarrow,\mathbf{k}+\mathbf{q}/2}^{\dagger}
c_{\uparrow,-\mathbf{k}+\mathbf{q}/2}^{\dagger}\big),
$
where $\mathbf q$ and $\mathbf k$ are the center-of-mass and relative momenta
of the pair, respectively. In terms of these operators, the interaction
becomes~\cite{vyasanakere12a}
\begin{equation}
\mathcal{H}_{\mathrm{int}}=-\frac{U}{2V}\sum_{\mathbf q \mathbf k \mathbf k'}
S_{\mathbf{k}}^{\mathbf q\dagger}S_{\mathbf{k}'}^{\mathbf q}.
\label{eq:interaction_hamiltonian_singlet}
\end{equation}
This form is useful because the helicity states have momentum-dependent
spin compositions. The interaction between two helicity states is therefore
determined by their overlap with the spin-singlet channel. Expressing the
interaction in terms of $S_{\mathbf k}^{\mathbf q}$ operators provides
the tool needed to formulate the two-body problem in the helicity basis,
as discussed next.

\subsection{Pair-size tensor for the two-body bound states}
\label{sec:twobody}

We first formulate the two-body bound-state problem for a general
center-of-mass momentum $\mathbf{q}$. Since the particle number is fixed to
two, this is a canonical problem in which the chemical potential does not
enter. The Hamiltonian $\mathcal H_0$ remains translationally invariant,
so the total momentum $\mathbf q$ is a conserved quantum number. However,
the center-of-mass and relative motions generally do not separate, since
the single-particle SOC energies depend separately on $\mathbf k+\mathbf q/2$
and $-\mathbf k+\mathbf q/2$. Thus, although a fixed-$\mathbf{q}$ sector is
well defined, the internal wavefunction generally depends parametrically
on the center-of-mass momentum.

To express the contact interaction in the helicity-pair basis, we use the
spin-singlet pair operator introduced in Sec.~\ref{sec:single_particle}.
We define the overlap of a helicity-pair state with the singlet channel
as
$
\mathcal{A}^\mathbf{q}_{ss'\mathbf{k}}=\langle S^\mathbf{q}_{\mathbf{k}}|\mathbf{q}\mathbf{k};ss'\rangle,
$
where
$
|S^\mathbf{q}_{\mathbf{k}} \rangle = S^{\mathbf{q}\dagger}_{\mathbf{k}} |0 \rangle
$
with $|0 \rangle$ the vacuum, and
$
|\mathbf{q}\mathbf{k};ss'\rangle=c_{s,\mathbf{k}+\mathbf{q}/2}^{\dagger}c_{s',-\mathbf{k}+\mathbf{q}/2}^{\dagger}|0\rangle
$
denotes a helicity-pair state~\cite{vyasanakere12a}.
More explicitly,
\begin{align}
\mathcal{A}^\mathbf{q}_{ss'\mathbf{k}} = \frac{1}{\sqrt{2}} \big[
u_{s,\mathbf{k}+\mathbf{q}/2}^{\uparrow}
u_{s',-\mathbf{k}+\mathbf{q}/2}^{\downarrow}
-
u_{s,\mathbf{k}+\mathbf{q}/2}^{\downarrow}
u_{s',-\mathbf{k}+\mathbf{q}/2}^{\uparrow}
\big],
\label{eq:singlet_overlap}
\end{align}
which satisfies
$
\mathcal{A}^{\mathbf q}_{ss'\mathbf{k}}
=-\mathcal{A}^{\mathbf q}_{s's,-\mathbf{k}}
$
due to fermionic antisymmetry.
For the two-body wavefunction, we restrict the relative-momentum sum to an
independent set of momenta, denoted by $\sum_{\mathbf{k}}'$, which contains
one representative from each pair $\mathbf{k}$ and $-\mathbf{k}$, i.e.,
half of the $\mathbf{k}$ space. In this restricted helicity-pair basis,
the most general two-body state with center-of-mass momentum $\mathbf{q}$ is
\begin{align}
|\Psi_{\mathbf q}\rangle=\sum_{s s' \mathbf{k}}'
\phi^{\mathbf q}_{s s' \mathbf{k}} |\mathbf{q}\mathbf{k}; s s'\rangle.
\end{align}
The coefficients $\phi^{\mathbf q}_{s s' \mathbf{k}}$ are wavefunction
amplitudes in the helicity-pair basis, and they satisfy the fermionic
antisymmetry condition
$
\phi^{\mathbf q}_{s s' \mathbf{k}} = - \phi^{\mathbf q}_{s' s, -\mathbf{k}}
$
when extended to the full relative-momentum space. The normalization is
fixed by
$
\sum_{s s' \mathbf{k}}'\left|\phi^{\mathbf q}_{s s' \mathbf{k}}\right|^2 = 1.
$

Substituting this restricted two-body ansatz into the Schrödinger equation
$(\mathcal H_0+\mathcal H_{\mathrm{int}})|\Psi_{\mathbf q}\rangle=E_{\mathbf q}|\Psi_{\mathbf q}\rangle$
and projecting onto the helicity-pair state $\langle \mathbf{q}\mathbf{k};s_1s_2|,$
with $\mathbf{k}$ in the primed set, gives
\begin{align}
\label{eq:projected_schrodinger}
\big(\varepsilon_{s_1,\mathbf{k}+\mathbf q/2}
+ &\varepsilon_{s_2,-\mathbf{k}+\mathbf q/2} - E_{\mathbf q} \big)
\phi^{\mathbf q}_{s_1s_2\mathbf{k}}
\\ &=
\frac{2U}{V}
\mathcal{A}^{\mathbf q *}_{s_1s_2\mathbf{k}}
\sum_{s_3s_4 \mathbf{k}'}'
\mathcal{A}^{\mathbf q}_{s_3s_4\mathbf{k}'}
\phi^{\mathbf q}_{s_3s_4\mathbf{k}'} .
\nonumber
\end{align}
The prefactor $2U/V$ in Eq.~\eqref{eq:projected_schrodinger} arises because
the singlet sums in Eq.~\eqref{eq:interaction_hamiltonian_singlet} are
unrestricted, while the two-body ansatz uses the primed momentum sum;
restricting both $S_{\mathbf k}^{\mathbf q\dagger}$ and
$S_{\mathbf k'}^{\mathbf q}$ each contributes a factor of two,
taking $-U/(2V)$ in Eq.~\eqref{eq:interaction_hamiltonian_singlet} to $-2U/V$.
To evaluate the singlet contribution in Eq.~\eqref{eq:projected_schrodinger},
we use the singlet weight
$
|\mathcal{A}^{\mathbf q}_{s s' \mathbf k}|^2 = \frac{1}{4}
(1 - s s' \hat{\mathbf d}_{\mathbf{k}+\mathbf q/2}
\cdot\hat{\mathbf d}_{-\mathbf{k}+\mathbf q/2}).
$
For the pair size, we focus on the zero-center-of-mass bound state,
setting $\mathbf{q}=\mathbf 0$ and denoting its energy by $E_{\mathbf 0}$,
in which case the singlet weight reduces to
$
|\mathcal{A}^\mathbf 0_{s s' \mathbf k}|^2=\frac{1}{4}
(1 - s s' \hat{\mathbf d}_{\mathbf k} \cdot\hat{\mathbf d}_{-\mathbf k}).
$
Since the SOC fields considered here are odd,
$
\mathbf d_{-\mathbf k}=-\mathbf d_{\mathbf k},
$
so that this further simplifies to
$
|\mathcal{A}^\mathbf{0}_{s s' \mathbf k}|^2=\frac14(1 + s s').
$
Consequently, only intra-helicity pair states carry nonzero singlet weight,
$
|\mathcal{A}^\mathbf{0}_{ss\mathbf k}|^2=\frac12,
$
and the $\mathbf q=\mathbf 0$ bound state lies entirely in the intra-helicity sector,
$
|\Psi_\mathbf{0}\rangle=\sum_{s\mathbf{k}}' \phi^{\mathbf 0}_{ss\mathbf{k}}
c_{s\mathbf{k}}^{\dagger} c_{s,-\mathbf{k}}^{\dagger} |0\rangle.
$
Using this projected ansatz in the $\mathbf{q}=\mathbf{0}$ form of
Eq.~\eqref{eq:projected_schrodinger} gives
\begin{equation}
\big(2\varepsilon_{s\mathbf{k}}-E_\mathbf{0}\big)\phi^{\mathbf 0}_{ss\mathbf{k}}
=\frac{2U}{V}\mathcal{A}^{\mathbf 0 *}_{ss\mathbf{k}}\sum_{s' \mathbf{k}'}'
\mathcal{A}^{\mathbf 0}_{s's'\mathbf{k}'}\phi^{\mathbf 0}_{s's'\mathbf{k}'}.
\label{eq:reduced_schrodinger}
\end{equation}
Solving Eq.~\eqref{eq:reduced_schrodinger} gives
$
\phi^{\mathbf 0}_{ss\mathbf{k}} = C \mathcal{A}^{\mathbf 0 *}_{ss\mathbf{k}}/
(2\varepsilon_{s\mathbf{k}}-E_\mathbf{0}),
$
where the constant $C$ is fixed by normalization,
$
|C|^2=2\big/\sum_{s\mathbf{k}}'1/(2\varepsilon_{s\mathbf{k}}-E_\mathbf{0})^2.
$

Requiring a nonzero bound-state amplitude then gives
$
1=(2U/V) \sum_{s\mathbf{k}}'
|\mathcal{A}^\mathbf{0}_{ss\mathbf{k}}|^2/
(2\varepsilon_{s\mathbf{k}}-E_\mathbf{0}),
$
which, using $|\mathcal{A}^\mathbf{0}_{ss\mathbf{k}}|^2=1/2$, reduces to the
bound-state equation
\begin{align}
1 = \frac{U}{V} \sum_{s\mathbf{k}}'
\frac{1}{2\varepsilon_{s\mathbf{k}}-E_\mathbf{0}}.
\label{eq:E0}
\end{align}
For a given $U$ and $\alpha$, this equation determines $E_\mathbf{0}$, consistent
with earlier results in the
literature~\cite{cappelluti07, Vyasanakere11, vyasanakere12a, yu2011socfeshbach}.
We refer to this two-body bound state, with binding energy $-E_\mathbf{0}$,
as a ``rashbon''~\cite{vyasanakere12}. In the strong-SOC limit,
the noninteracting density of states diverges on the ring or sphere of
$-$-helicity band minima, so the momentum sum in Eq.~\eqref{eq:E0}
diverges as $\alpha\to\infty$ at fixed $E_\mathbf{0}$; a nontrivial solution $E_\mathbf{0}$
therefore exists for arbitrarily weak attraction, $U\to0^+$. This is the
origin of the interaction-independent strong-SOC asymptotes reported in
Sec.~\ref{sec:numerical_results}: for any $U\neq0$, sufficiently strong
SOC always binds a ``rashbon'', driving the system into a
weakly-interacting ``rashbon'' BEC.

We characterize the size of the two-body bound state by the square root of
the trace of the two-body localization tensor of the spin-basis wavefunction
$\Psi_{\sigma\sigma'}(\mathbf r)$, i.e.,
$
\bar{\xi}_{\mathrm{2b}} = \sqrt{\mathrm{Tr} (\boldsymbol{\xi}^2_{\mathrm{2b}})},
$
where~\cite{iskin25}
\begin{equation}
(\xi^2_{\mathrm{2b}})_{ij}=\frac{\sum_{\sigma \sigma'} \int d^d\mathbf{r} \, r_i r_j
|\Psi_{\sigma\sigma'}(\mathbf{r})|^2}{\sum_{\sigma \sigma'}\int d^d \mathbf{r}
|\Psi_{\sigma\sigma'}(\mathbf{r})|^2}.
\label{eq:two_body_localization_tensor}
\end{equation}
Here $r_i$ with $i \in \{x,y,z\}$ denotes the $i$th component of the relative
coordinate $\mathbf{r}=\mathbf{r}_1-\mathbf{r}_2$. Since the two-body state
was expanded over a restricted set of relative momenta, the corresponding
spin-basis wavefunction is obtained by antisymmetrizing the contribution
from each representative momentum,
$
\Psi_{\sigma\sigma'}(\mathbf{r}) = \frac{1}{V} \sum_{s \mathbf{k}}'
\phi^{\mathbf 0}_{ss\mathbf{k}}
(e^{i\mathbf{k}\cdot\mathbf{r}} u_{s\mathbf{k}}^\sigma u_{s,-\mathbf{k}}^{\sigma'}
-e^{-i\mathbf{k}\cdot\mathbf{r}} u_{s\mathbf{k}}^{\sigma'} u_{s,-\mathbf{k}}^\sigma).
$
This construction satisfies the fermionic antisymmetry condition
$
\Psi_{\sigma\sigma'}(\mathbf r)=-\Psi_{\sigma'\sigma}(-\mathbf r).
$
For momenta included in the restricted sum, it is useful to define
$
F_{\sigma\sigma'}(\mathbf k) = \sum_s \phi^{\mathbf 0}_{ss\mathbf k}
w_{s\mathbf k }^{\sigma\sigma'},
$
where
$
w_{s\mathbf k}^{\sigma\sigma'}=u_{s\mathbf k}^\sigma u_{s,-\mathbf k}^{\sigma'}.
$
By converting the real-space coordinate factors to momentum derivatives
in the Fourier representation, and using integration by parts,
Eq.~\eqref{eq:two_body_localization_tensor}
can be expressed in momentum space as
\begin{equation}
(\xi^2_{\mathrm{2b}})_{ij}=\frac{\sum_{\sigma \sigma' \mathbf{k}}'
\partial_i F_{\sigma\sigma'}(\mathbf{k})
\partial_j F_{\sigma\sigma'}^{*}(\mathbf{k})}
{\sum_{\sigma \sigma' \mathbf{k}}'
|F_{\sigma\sigma'}(\mathbf{k})|^2},
\qquad (\mathbf{q} = \mathbf{0}).
\label{eq:two_body_localization_momentum}
\end{equation}
Although the helicity spinors are defined up to momentum-dependent phases,
the combinations
$
\phi^{\mathbf 0}_{ss \mathbf k}w_{s\mathbf k}^{\sigma\sigma'}
$
entering $F_{\sigma\sigma'}(\mathbf k)$ are gauge independent.
Indeed, $w_{s\mathbf k }^{\sigma\sigma'}$ and $\mathcal{A}^{\mathbf 0}_{ss\mathbf k}$
acquire the same gauge phase, while
$
\phi^{\mathbf 0}_{ss \mathbf k}\propto \mathcal{A}^{\mathbf 0 *}_{ss\mathbf k}
$
acquires the opposite phase, so the two cancel. Thus,
Eq.~\eqref{eq:two_body_localization_momentum} is gauge invariant.

To specify the helicity spinors, we parametrize the direction of the SOC field as
$
\hat{\mathbf d}_{\mathbf k}= (\sin\theta_{\mathbf k}\cos\varphi_{\mathbf k},
\sin\theta_{\mathbf k}\sin\varphi_{\mathbf k},\cos\theta_{\mathbf k} ),
$
where
$
0\leq\theta_{\mathbf k}\leq\pi
$
and
$
0\leq\varphi_{\mathbf k}<2\pi
$
are the polar and azimuthal angles of $\hat{\mathbf d}_{\mathbf k}$,
respectively. For the 2D and 3D Rashba SOC fields, $\theta_{\mathbf k}=\pi/2$.
Since the SOC fields considered here are odd under momentum inversion,
$
\hat{\mathbf d}_{-\mathbf k}=-\hat{\mathbf d}_{\mathbf k},
$
and the corresponding angular variables satisfy
$
\theta_{-\mathbf k}=\pi-\theta_{\mathbf k}
$
and
$
\varphi_{-\mathbf k}=\varphi_{\mathbf k}+\pi
$
modulo $2\pi$.
For every momentum $\mathbf k$ included in the primed sum, we choose
$
|u_{+,\mathbf k} \rangle \equiv [\cos(\theta_{\mathbf k}/2), \,
e^{i\varphi_{\mathbf k}}\sin(\theta_{\mathbf k}/2)]^\mathrm{T}
$
and
$
|u_{-,\mathbf k} \rangle \equiv [\sin(\theta_{\mathbf k}/2), \,
-e^{i\varphi_{\mathbf k}}\cos(\theta_{\mathbf k}/2)]^\mathrm{T}.
$
For their partners at $-\mathbf k$, we fix them by our time-reversal
gauge convention
$
|u_{s,-\mathbf k} \rangle \equiv i\sigma_y |u_{s \mathbf k} \rangle^{*},
$
which gives
$
|u_{+,-\mathbf k} \rangle \equiv [e^{-i\varphi_{\mathbf k}}\sin(\theta_{\mathbf k}/2), \,
-\cos(\theta_{\mathbf k}/2)]^\mathrm{T}
$
and
$
|u_{-,-\mathbf k} \rangle \equiv [-e^{-i\varphi_{\mathbf k}}\cos(\theta_{\mathbf k}/2), \,
-\sin(\theta_{\mathbf k}/2)]^\mathrm{T}.
$
For the ordered pair $(\mathbf k,-\mathbf k)$, these spinors give
$
\mathcal{A}^{\mathbf0}_{ss\mathbf k}=-1/\sqrt2
$
and
$
\partial_i \mathcal{A}^{\mathbf0}_{ss\mathbf k}=0,
$
while the partner momentum $-\mathbf k$, which is not included
independently in the primed sum, satisfies
$
\mathcal{A}^{\mathbf0}_{ss,-\mathbf k}=
-\mathcal{A}^{\mathbf0}_{ss\mathbf k}=1/\sqrt2.
$

The helicity-pair spinors satisfy
$
\sum_{\sigma \sigma'} w_{s\mathbf k}^{\sigma\sigma' *}
w_{s'\mathbf k}^{\sigma\sigma'}=\delta_{ss'},
$
where $\delta_{ss'}$ is a Kronecker delta.
Using this orthonormality relation together with the quantum metric
definitions introduced in Sec.~\ref{sec:single_particle},
Eq.~\eqref{eq:two_body_localization_momentum} can be written as
$
(\xi^2_{\mathrm{2b}})_{ij}= \sum_{s\mathbf{k}}' [
\partial_i \phi^{\mathbf 0 *}_{ss\mathbf{k}}\partial_j \phi^{\mathbf 0}_{ss\mathbf{k}}
+ \phi^{\mathbf 0 *}_{ss\mathbf{k}}(\phi^{\mathbf 0}_{ss\mathbf{k}} - \phi^{\mathbf 0}_{-s,-s, \mathbf{k}})
g^{\mathbf{k}}_{ij} ]
/ \sum_{s\mathbf{k}}'|\phi^{\mathbf 0}_{ss\mathbf{k}}|^2.
$
Substituting the bound-state solution of Eq.~\eqref{eq:reduced_schrodinger},
the overall constant cancels and the localization tensor decomposes as
$
(\xi^2_{\mathrm{2b}})_{ij}
=
(\xi^2_{\mathrm{2b}})^\mathrm{intra}_{ij}
+
(\xi^2_{\mathrm{2b}})^\mathrm{inter}_{ij},
$
where
\begin{align}
(\xi^2_{\mathrm{2b}})^\mathrm{intra}_{ij} & =
\frac{ 4\sum_{s\mathbf{k}}'
\frac{\partial_i\varepsilon_{s\mathbf{k}} \partial_j\varepsilon_{s\mathbf{k}}}
{(2\varepsilon_{s\mathbf{k}}-E_\mathbf{0})^4}}
{\sum_{s\mathbf{k}}' \frac{1}{(2\varepsilon_{s\mathbf{k}}-E_\mathbf{0})^2}},
\label{eq:two_body_pair_size_intra}
\\
(\xi^2_{\mathrm{2b}})_{ij}^\mathrm{inter} & =
\frac{\sum_{s\mathbf{k}}'\big[
\frac{1}{(2\varepsilon_{s\mathbf{k}}-E_\mathbf{0})^2}
-
\frac{1}{(2\varepsilon_{s\mathbf{k}}-E_\mathbf{0})
(2\varepsilon_{-s,\mathbf{k}}-E_\mathbf{0})
}
\big]
g^{\mathbf{k}}_{ij}}
{\sum_{s\mathbf{k}}'
\frac{1}{(2\varepsilon_{s\mathbf{k}}-E_\mathbf{0})^2}}.
\label{eq:two_body_pair_size_inter}
\end{align}
We refer to these two terms as the intraband and interband contributions,
respectively, and use this terminology throughout: the intraband
contribution is determined by the helicity-band structure, whereas the
interband contribution contains the quantum metric and is accordingly
referred to as the quantum-geometric contribution. In the absence of SOC,
Eq.~\eqref{eq:two_body_pair_size_intra} reproduces the textbook result for
the usual two-body problem in vacuum~\cite{leggett}, while
Eq.~\eqref{eq:two_body_pair_size_inter} is caused by SOC and originates
from virtual interband transitions between the helicity bands. Having
established the intraband-interband decomposition for the exact two-body
bound states, we now extend the same framework to the many-body Cooper pair
within mean-field BCS theory at $T=0$.

\subsection{Pair-size tensor for the Cooper pairs at $T = 0$}
\label{sec:manybody}

We now turn to the average size of Cooper pairs within the mean-field
BCS theory at zero temperature, assuming time-reversal symmetry and the
same spinor convention as in the two-body problem.
Motivated by the pair-size calculations in the usual BCS-BEC crossover
theories with no SOC~\cite{sademelo93, pistolesi94, engelbrecht97, schunck08},
here we extend this concept to its tensorial form to highlight its deeper
connection to the quantum-metric tensor of the helicity states in the
presence of SOC.

For this purpose, we define the uniform pairing amplitude as
$
\Delta_0 = \frac{U}{V}\sum_{\mathbf{k}}
\left\langle \mathrm{BCS}\right| c_{\downarrow, -\mathbf{k}}
c_{\uparrow \mathbf{k}}\left|\mathrm{BCS}\right\rangle,
$
which we choose to be real and positive. In the helicity basis, the BCS
ground state for the stationary pairs
reads~\cite{yu2011socfeshbach, yu2014coherence}
\begin{align}
|\mathrm{BCS}\rangle = \prod_{s\mathbf{k}}' (\mathcal{U}_{s\mathbf{k}}
+ \mathcal{V}_{s\mathbf{k}} c_{s\mathbf{k}}^{\dagger}
c_{s,-\mathbf{k}}^{\dagger})|0\rangle,
\end{align}
where the primed product runs over the same independent set of momenta
used for the two-body problem, and
$
E_{s\mathbf{k}}=\sqrt{\xi_{s\mathbf{k}}^2+|\Delta_s(\mathbf{k})|^2}
$
is the quasiparticle spectrum with helicity-resolved gap
$
\Delta_s(\mathbf{k})=\sqrt{2}\Delta_0 \mathcal{A}_{ss\mathbf{k}}^{\mathbf0*}.
$
With $\mathcal U_{s\mathbf k}$ chosen real and positive, the coherence
factors are
$
\mathcal{U}_{s\mathbf{k}}^2 = \frac{1}{2}\big(1+\frac{\xi_{s\mathbf{k}}}
{E_{s\mathbf{k}}}\big),
$
$
|\mathcal{V}_{s\mathbf{k}}|^2 = \frac{1}{2}\big(1-\frac{\xi_{s\mathbf{k}}}
{E_{s\mathbf{k}}}\big),
$
and
$
\mathcal U_{s\mathbf k}^{*}\mathcal V_{s\mathbf k}
= \frac{\Delta_s(\mathbf k)}{2E_{s\mathbf k}},
$
where $|\mathcal{V}_{s\mathbf{k}}|^2$ is the occupation probability of the
helicity state $|u_{s\mathbf k} \rangle$ and
$\mathcal{U}^{*}_{s\mathbf{k}}\mathcal{V}_{s\mathbf{k}}$ is the associated
anomalous amplitude.
For every representative momentum $\mathbf k$ in the primed set,
$
\mathcal{A}_{ss\mathbf{k}}^{\mathbf0}=-1/\sqrt2,
$
so that $\Delta_s(\mathbf{k})=-\Delta_0$ for both helicity branches; with
$\mathcal U_{s\mathbf k} \ge 0$, this fixes
$\mathcal V_{s\mathbf k}=-|\mathcal V_{s\mathbf k}|$ over the primed set.
At the partner momentum $-\mathbf{k}$,
$
\mathcal{A}_{ss,-\mathbf{k}}^{\mathbf0}=
-\mathcal{A}_{ss\mathbf{k}}^{\mathbf0}=1/\sqrt2,
$
giving $\Delta_s(-\mathbf{k})=\Delta_0=-\Delta_s(\mathbf{k})$. The
intra-helicity pairing amplitude is thus odd under
$\mathbf k \rightarrow -\mathbf k$, as required for pairing between two
fermions in the same helicity branch, while its magnitude stays constant,
$
|\Delta_s(\mathbf{k})|=\Delta_0,
$
so that the quasiparticle spectrum simplifies to
$
E_{s\mathbf{k}}=\sqrt{\xi_{s\mathbf{k}}^2+\Delta_0^2}.
$
Here, $\Delta_0$ and $\mu$ are fixed self-consistently by demanding that
the mean-field gap and number operators reproduce their assumed values
in the BCS ground state, i.e., by the gap equation
$
\Delta_0 = \frac{U}{V}\sum_{\mathbf{k}}
\left\langle \mathrm{BCS}\right| c_{\downarrow, -\mathbf{k}}
c_{\uparrow \mathbf{k}}\left|\mathrm{BCS}\right\rangle,
$
and the number equation
$
n = \frac{1}{V} \sum_{\sigma \mathbf{k}}
\left\langle \mathrm{BCS}\right| c_{\sigma \mathbf{k}}^\dagger
c_{\sigma \mathbf{k}}\left|\mathrm{BCS}\right\rangle.
$
Evaluated in the helicity basis, these reduce to
\begin{align}
\label{eq:op}
\frac{1}{U} &= \frac{1}{V} \sum_{s\mathbf{k}}
\frac{|\mathcal{A}^{\mathbf 0}_{ss\mathbf{k}}|^2}{2E_{s\mathbf{k}}}
= \frac{1}{2V} \sum_{s\mathbf{k}} \frac{1}{2E_{s\mathbf{k}}},
\\
\label{eq:ne}
n &=
\frac{1}{V} \sum_{s\mathbf{k}} \frac{1}{2}
\Big( 1-\frac{\xi_{s\mathbf{k}}}{E_{s\mathbf{k}}} \Big),
\end{align}
which suffice to determine the BCS-BEC crossover physics at
$T=0$~\cite{engelbrecht97, yu2011socfeshbach, yu2014coherence,vyasanakere12a,jiang11,he2012rashba2d,he2012weyl, seo12}.
As a consistency check, Eq.~\eqref{eq:op}
reduces to the two-body bound-state equation~\eqref{eq:E0} in the dilute
limit $\Delta_0\to0$, where $\mu\to E_\mathbf{0}/2$.

We note in passing that $\mathcal{V}_{s\mathbf{k}}$ carries no explicit
momentum-dependent phase in our BCS ground state; this follows from the
time-reversal convention adopted here, which fixes the overall phase
relating the helicity states at $\mathbf{k}$ and $-\mathbf{k}$ and renders
their singlet overlap real and momentum-independent over the primed set.
In the helicity conventions of Refs.~\cite{yu2011socfeshbach,
yu2014coherence, seo12}, by contrast, the helicity-resolved order parameter
$\Delta_s(\mathbf k)$ carries an explicit momentum-dependent phase
$\varphi_\mathbf{k}$, which also enters the intra-helicity pairing term;
over the full momentum space, this phase reproduces the same odd-parity
relation $\Delta_s(-\mathbf k)=-\Delta_s(\mathbf k)$ obtained above.
Because the pair size is gauge invariant, our convention simplifies its
evaluation: momentum derivatives of the anomalous amplitude act only on
$E_{s\mathbf k}$, without generating additional derivatives from a
pairing phase. Once the difference in helicity-basis conventions is
accounted for, both formulations describe the same BCS state in the spin
basis.

We define the Cooper-pair wavefunction from the BCS ground state as
$
\Phi_{\sigma\sigma'}(\mathbf{r})=\left\langle \mathrm{BCS}\right|
\psi_{\sigma}(\mathbf{r})\psi_{\sigma'}(\mathbf{0})
\left|\mathrm{BCS}\right\rangle
$
~\cite{yu2011socfeshbach, yu2014coherence}.
In direct analogy with the two-body problem, we characterize the average
size of Cooper pairs by the square root of the trace of the Cooper-pair
localization tensor, i.e.,
$
\bar{\xi}_\mathrm{Cp} = \sqrt{\mathrm{Tr}(\boldsymbol{\xi}^2_{\mathrm{Cp}})},
$
where
\begin{equation}
(\xi^2_{\mathrm{Cp}})_{ij} = \frac{\sum_{\sigma \sigma'}
\int d^d \mathbf{r}\,r_i r_j|\Phi_{\sigma\sigma'}(\mathbf{r})|^2}
{\sum_{\sigma \sigma'} \int d^d \mathbf{r}\,|\Phi_{\sigma\sigma'}(\mathbf{r})|^2}.
\label{eq:cooper_pair_localization_tensor}
\end{equation}
Converting the real-space coordinate factors to momentum derivatives
in the Fourier representation, and using integration by parts exactly
as was done for the two-body case,
Eq.~\eqref{eq:cooper_pair_localization_tensor} can be expressed in
momentum space as
$
(\xi^2_{\mathrm{Cp}})_{ij} = \sum_{s\mathbf{k}}' [
\partial_i (\mathcal{U}^{*}_{s\mathbf{k}}\mathcal{V}_{s\mathbf{k}})
\partial_j (\mathcal{U}_{s\mathbf{k}}\mathcal{V}^{*}_{s\mathbf{k}})
+ \mathcal{U}^{*}_{s\mathbf{k}}\mathcal{V}_{s\mathbf{k}}
(\mathcal{U}_{s\mathbf{k}}\mathcal{V}_{s\mathbf{k}}^*
- \mathcal{U}_{-s,\mathbf{k}}\mathcal{V}^{*}_{-s,\mathbf{k}})
g^{\mathbf{k}}_{ij} ]
/ \sum_{s\mathbf{k}}'|\mathcal{U}^{*}_{s\mathbf{k}}\mathcal{V}_{s\mathbf{k}}|^2.
$
This expression has the same structure as the two-body result,
with the bound-state amplitude $\phi^{\mathbf 0}_{ss\mathbf{k}}$ replaced by the
BCS anomalous amplitude
$
\mathcal{U}^{*}_{s\mathbf{k}}\mathcal{V}_{s\mathbf{k}}.
$
Substituting the BCS anomalous amplitude for the momenta included in the
primed sums, the overall constant again cancels and the localization tensor
decomposes as
$
(\xi^2_{\mathrm{Cp}})_{ij}
=
(\xi^2_{\mathrm{Cp}})^\mathrm{intra}_{ij}
+
(\xi^2_{\mathrm{Cp}})^\mathrm{inter}_{ij},
$
into the analogous intraband and interband contributions defined for the
two-body problem. The resulting intraband and interband contributions are
\begin{align}
(\xi^2_{\mathrm{Cp}})^\mathrm{intra}_{ij}
&= \frac{ \sum_{s\mathbf{k}}'
\frac{\xi_{s\mathbf{k}}^{2}} {E_{s\mathbf{k}}^{6}}
\partial_i\xi_{s\mathbf{k}} \partial_j\xi_{s\mathbf{k}} }
{\sum_{s\mathbf{k}}'\frac{1}{E_{s\mathbf{k}}^{2}}},
\label{eq:many_body_pair_size_intra}
\\
(\xi^2_{\mathrm{Cp}})^\mathrm{inter}_{ij} &=
\frac{\sum_{s\mathbf{k}}'\big( \frac{1}{E_{s\mathbf{k}}^{2}}
- \frac{1}{E_{s\mathbf{k}}E_{-s,\mathbf{k}}} \big)
g^{\mathbf{k}}_{ij}}
{\sum_{s\mathbf{k}}'\frac{1}{E_{s\mathbf{k}}^{2}}}.
\label{eq:many_body_pair_size_inter}
\end{align}
It is reassuring that these expressions recover the two-body results
Eqs.~\eqref{eq:two_body_pair_size_intra}
and~\eqref{eq:two_body_pair_size_inter} in the dilute limit, and that,
in the absence of SOC, Eq.~\eqref{eq:many_body_pair_size_intra}
reproduces the well-known result for the usual BCS-BEC crossover
problem~\cite{sademelo93, pistolesi94, engelbrecht97, schunck08}.

At the mean-field level, $\xi_\mathrm{Cp}$ coincides with the quantity
called the coherence length in Ref.~\cite{yu2014coherence}; it should
not be confused with the coherence length $\xi_0$ defined below from
the Gaussian fluctuations of the order parameter
(see Sec.~\ref{sec:literature}).

\subsection{Coherence-length tensor at $T = 0$}
\label{sec:coherencelength}

To compare with the pair size, here we derive a zero-temperature
coherence-length tensor from the effective Gaussian action for the
order-parameter fluctuations~\cite{pistolesi96, iskin24c, elden26},
obtained by expanding the effective action around the uniform mean-field
saddle point. We write the pairing field as
$
\Delta_Q = \Delta_0\,\delta_{Q0}+\Lambda_Q,
$
where $\Lambda_Q$ is a complex fluctuation field and $Q=(\mathbf q,i\nu_\ell)$
is a collective four-momentum, with $\nu_\ell=2\pi\ell T$ the bosonic
Matsubara frequency and $\ell\in\mathbb Z$.

Using the Grassmann functional-integral formalism, one can show that
the effective action is quadratic in $\Lambda_Q$ to leading
order~\cite{pistolesi96, engelbrecht97, jiang11, vyasanakere12a, he2012weyl,
he2012rashba2d, iskin20a},
\begin{equation}
\mathcal{S}_{\mathrm{eff}}^{(2)}=\frac{1}{2}\sum_Q
\boldsymbol{\Lambda}_Q^{\dagger} \mathbf M(Q)
\boldsymbol{\Lambda}_Q,
\label{eq:gaussian_action}
\end{equation}
where
$
\boldsymbol{\Lambda}_Q = (\Lambda_Q, \, \Lambda_{-Q}^{*})^{\mathrm T}
$
is the fluctuation spinor, and $\mathbf M(Q)$ is the inverse
fluctuation propagator. The coherence length is obtained below from the
long-wavelength, static expansion of this matrix.
Using the short-hand notations,
$
E_s^\pm \equiv E_{s,\pm\mathbf{k}+\mathbf q/2},
$
$
\xi_s^\pm \equiv \xi_{s,\pm\mathbf{k}+\mathbf q/2}
$
and
$
\hat{\mathbf d}^{\pm} \equiv \hat{\mathbf d}_{\pm\mathbf{k}+\mathbf q/2},
$
the matrix elements at $T=0$ can be written as
\begin{align}
\label{eq:fluctuation_M11}
M_{11}(Q)
&= \frac{V}{U}
+\frac{1}{16}\sum_{ss'\mathbf{k}}
\frac{1-ss'\,\hat{\mathbf d}^{+}\cdot\hat{\mathbf d}^{-}}
{E_s^+E_{s'}^-}
\\
&\times \Big[
\frac{(E_s^++\xi_s^+)(E_{s'}^-+\xi_{s'}^-)}
{i\nu_\ell-E_s^+-E_{s'}^-}
-\frac{(E_s^+-\xi_s^+)(E_{s'}^--\xi_{s'}^-)}
{i\nu_\ell+E_s^++E_{s'}^-}
\Big],
\nonumber  \\
M_{12}(Q)
&=\frac{\Delta_0^2}{16}\sum_{ss'\mathbf{k}}
\frac{1-ss'\,\hat{\mathbf d}^{+}\cdot\hat{\mathbf d}^{-}}
{E_s^+E_{s'}^-}
\nonumber  \\
& \times \Big(
\frac{1}{i\nu_\ell+E_s^++E_{s'}^-}
-\frac{1}{i\nu_\ell-E_s^+-E_{s'}^-}
\Big).
\label{eq:fluctuation_M12}
\end{align}
The remaining elements satisfy
$
M_{22}(Q)=M_{11}(-Q)$ and $M_{21}(Q)=M_{12}(Q).
$
In the static limit, the amplitude fluctuations are governed by the
combination
$
M_{11}(\mathbf q,0)+M_{12}(\mathbf q,0).
$
Its expansion for $\mathbf q\to \mathbf{0}$ takes the form
$
M_{11}(\mathbf q,0)+M_{12}(\mathbf q,0)=A+\sum_{ij}C_{ij}q_iq_j+\mathcal O(\mathbf{q}^4).
$

The gradient coefficient decomposes into intraband and interband
contributions,
$
C_{ij}=C_{ij}^{\mathrm{intra}}+C_{ij}^{\mathrm{inter}}.
$
For the spin-orbit-coupled Fermi gas considered here, the static
coefficient and the two contributions to the gradient coefficient are
given by
\begin{align}
A&=\sum_{s\mathbf k }\frac{\Delta_0^2}{4E_{s\mathbf k}^{3}},
\label{eq:coherence_static_coefficient}
\\
C_{ij}^{\mathrm{intra}} & = \sum_{s\mathbf k}
\frac{1}{16E_{s\mathbf k}^{3}}
\Big(1-\frac{5\Delta_0^2\xi_{s\mathbf k}^{2}}{E_{s\mathbf k}^{4}}\Big)
\partial_i\xi_{s\mathbf k}\,\partial_j\xi_{s\mathbf k},
\label{eq:coherence_gradient_intra}
\\
C_{ij}^{\mathrm{inter}} & =\sum_{s\mathbf{k}}\Big[
\frac{\left(\xi_{s\mathbf{k}}-\xi_{-s,\mathbf{k}}\right)^2+4\Delta_0^2}
{16E_{s\mathbf{k}}E_{-s,\mathbf{k}}(E_{s\mathbf{k}}+E_{-s,\mathbf{k}})}
-\frac{\Delta_0^2}{8E_{s\mathbf{k}}^3}\Big]
g^{\mathbf{k}}_{ij}.
\label{eq:coherence_gradient_inter}
\end{align}
In the absence of SOC, Eqs.~\eqref{eq:coherence_static_coefficient}
and~\eqref{eq:coherence_gradient_intra} reproduce the well-known result
for the usual BCS-BEC crossover problem~\cite{engelbrecht97}.
We note that Eq.~\eqref{eq:coherence_gradient_inter} was stated incorrectly
in Ref.~\cite{iskin20a}; however, this error does not affect the velocity
of the low-energy collective modes reported there, since
$C_{ij}$ does not enter that calculation.
See Appendix~\ref{app:interband-gradient} for the derivation of
Eq.~\eqref{eq:coherence_gradient_inter}.
The coherence-length tensor is then defined as
$
(\xi_0^2)_{ij}=C_{ij}/A.
$
Accordingly, it too decomposes as
$
(\xi_0^2)_{ij}=(\xi_{0}^2)^\mathrm{intra}_{ij}+(\xi_{0}^2)^\mathrm{inter}_{ij},
$
where~\cite{iskin24c}
\begin{align}
(\xi_{0}^2)^\mathrm{intra}_{ij} = \frac{C_{ij}^{\mathrm{intra}}}{A},
\quad
(\xi_{0}^2)^\mathrm{inter}_{ij}=\frac{C_{ij}^{\mathrm{inter}}}{A}.
\label{eq:cl}
\end{align}
This decomposition follows the same intraband-interband distinction
introduced for the pair size in Secs.~\ref{sec:twobody} and~\ref{sec:manybody}.
The behavior of the intraband and quantum-geometric contributions to
both the pair size and coherence length for the SOC models considered
here is discussed in Sec.~\ref{sec:numerical_results}.

The coherence length $\xi_0$ characterizes the spatial recovery of the
order-parameter amplitude following a small, static local disturbance.
At each coupling, the static amplitude susceptibility takes the
long-wavelength form
$
\chi_{\mathrm{amp}}(\mathbf{q},0)\propto[A+\sum_{ij}C_{ij}q_iq_j]^{-1},
$
which has a characteristic decay length
$
\sqrt{(\xi_0^2)_{ii}}=\sqrt{C_{ii}/A}
$
along each principal axis, so that $\xi_0$ is the static amplitude
correlation length within the Gaussian approximation. Since
$A>0$ for all couplings at $T=0$, the static amplitude channel is
stable and this expansion is
well defined throughout the crossover, so the identification is not
restricted to the weak-coupling regime. Since the susceptibility
determines the amplitude change induced by the disturbance, the same
length governs how the amplitude heals toward its uniform equilibrium
value away from it. The phase channel, governed by $M_{11}-M_{12}$,
instead vanishes at zero momentum and frequency, reflecting the gapless
Goldstone mode, and its leading gradient term determines the phase
stiffness rather than a coherence length. In the weak-coupling BCS
limit, $\xi_0$ has the familiar BCS coherence-length scaling, while in
the BEC limit it is proportional to the Gross-Pitaevskii healing length
of the molecular condensate, as shown explicitly in
Sec.~\ref{sec:3d-rashba}. In the absence of SOC, $\xi_0$ is furthermore
known to coincide, up to a coefficient of order unity, with the
Ginzburg-Landau coherence length near the critical
temperature~\cite{pistolesi96}.

Unlike the pair-size tensor, for which the trace provides the natural
rotationally invariant measure, the coherence-length tensor encodes an
anisotropic stiffness, or equivalently an inverse-mass-like tensor, in
the real-space Ginzburg-Landau functional. Its natural scalar measure is
therefore the geometric mean of its principal-axis values,
$
\bar{\xi}_0 = [\det (\boldsymbol{\xi}_0^2)]^\frac{1}{2d},
$
in analogy with the anisotropic healing length of a neutral superfluid
with an anisotropic effective mass.

\section{Numerical results}
\label{sec:numerical_results}

In this section, our primary interest is in how the intraband and
quantum-geometric interband contributions to the many-body pair size
and coherence length vary across the interaction-SOC parameter
space at $T = 0$. We consider Fermi gases with 3D Rashba,
3D Weyl and 2D Rashba SOC at fixed particle density $n$. For each
interaction strength and SOC strength $\alpha$, the chemical potential
$\mu$ and pairing amplitude $\Delta_0$ are determined
self-consistently from the gap equation, Eq.~\eqref{eq:op}, and
the number equation, Eq.~\eqref{eq:ne}. These solutions are then used to
evaluate the pair-size tensor $(\xi_{\mathrm{Cp}}^2)_{ij}$ from
Eqs.~\eqref{eq:many_body_pair_size_intra}--\eqref{eq:many_body_pair_size_inter},
and the coherence-length tensor $(\xi_0^2)_{ij}$ from Eq.~\eqref{eq:cl}.
We use two complementary density conventions, matched to the dimensionality
of the system. In 3D, $n = k_\mathrm{F}^3/(3\pi^2)$ defines
the Fermi momentum $k_\mathrm{F}$, and we parametrize the interaction
strength by the inverse scattering length $1/(k_\mathrm{F}a_s)$.
In 2D, we instead use $n = k_\mathrm{F}^2/(2\pi)$, and parametrize the
interaction strength by the binding energy ratio $E_b/\varepsilon_\mathrm{F}$.
In both cases, $\varepsilon_\mathrm{F}=k_\mathrm{F}^2/(2m)$ is the Fermi energy,
and we report all results in terms of the dimensionless SOC strength
$m\alpha/k_\mathrm{F}$. In all three models the interband contributions
vanish in the zero-SOC limit and the intraband term remains the dominant
contribution to both length scales throughout the calculated parameter
range.

We note that corrections beyond the present mean-field and
Gaussian-fluctuation treatment may affect the positions and magnitudes
of the reported extrema, particularly in strongly interacting regimes
such as the vicinity of unitarity in three dimensions.

\begin{figure*}[htb]
\centering
\includegraphics[width=\textwidth]{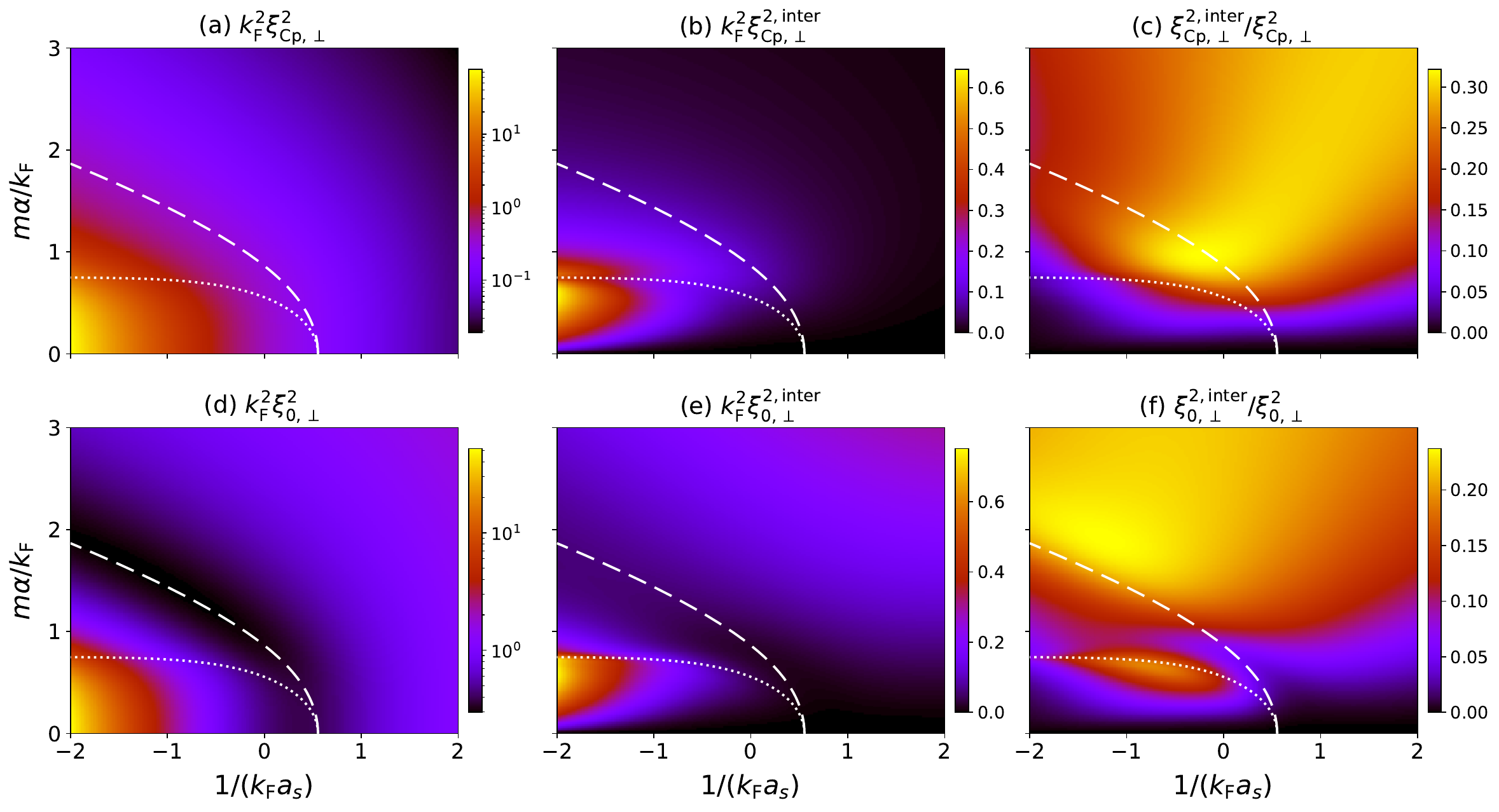}
\caption{
Transverse pair-size and coherence-length components for the
3D Rashba model as functions of $1/(k_\mathrm{F}a_s)$ and
$m\alpha/k_\mathrm{F}$.
Panels (a)--(c) show the total transverse pair size
$k_\mathrm{F}^2\xi_{\mathrm{Cp},\perp}^2$, its interband contribution
$k_\mathrm{F}^2\xi_{\mathrm{Cp},\perp}^{2,\mathrm{inter}}$,
and the relative interband contribution
$\xi_{\mathrm{Cp},\perp}^{2,\mathrm{inter}}/\xi_{\mathrm{Cp},\perp}^2$,
respectively. Panels (d)--(f) show the corresponding transverse quantities
for the coherence length, $k_\mathrm{F}^2\xi_{0,\perp}^2$,
$k_\mathrm{F}^2\xi_{0,\perp}^{2,\mathrm{inter}}$, and
$\xi_{0,\perp}^{2,\mathrm{inter}}/\xi_{0,\perp}^2$. Logarithmic color scales
are used for the total quantities, and linear scales for the interband
and relative interband contributions. White dashed and dotted contours
indicate $\mu=-m\alpha^2/2$ and $\mu=0$, respectively.
}
\label{fig:3d-rashba-lengthscales}
\end{figure*}
\subsection{Three-dimensional system with Rashba SOC}
\label{sec:3d-rashba}

For the 3D Rashba model, rotational symmetry about the $z$ axis implies
$
(\xi_{\mathrm{Cp}}^2)_{xx}=(\xi_{\mathrm{Cp}}^2)_{yy},
$
while $(\xi_{\mathrm{Cp}}^2)_{zz}$ is generally distinct at finite SOC.
The coherence-length tensor satisfies the analogous relations,
$
(\xi_0^2)_{xx}=(\xi_0^2)_{yy},
$
with $(\xi_0^2)_{zz}$ generally different from the transverse component.
All off-diagonal components of both tensors vanish, so we define the
scalar transverse components as
$
\xi_{\mathrm{Cp},\perp}^2\equiv(\xi_{\mathrm{Cp}}^2)_{xx}
=(\xi_{\mathrm{Cp}}^2)_{yy}
$
and
$
\xi_{0,\perp}^2\equiv(\xi_0^2)_{xx}=(\xi_0^2)_{yy},
$
and use the same notation separately for their intraband and interband
contributions, i.e., 
$
\xi_{\mathrm{Cp},\perp}^{2,\mathrm{inter}}\equiv(\xi_{\mathrm{Cp}}^2)^{\mathrm{inter}}_{xx},
$
and likewise for the other components.
Since the helicity states are independent of $k_z$, the quantum metric
has no $zz$ component; consequently, the longitudinal components are
purely intraband, whereas the transverse components contain both
intraband and interband contributions. We therefore present the
dimensionless transverse quantities $k_\mathrm{F}^2\xi_{\mathrm{Cp},\perp}^2$
and $k_\mathrm{F}^2\xi_{0,\perp}^2$, together with their quantum-geometric
interband contributions and relative interband fractions.

Figure~\ref{fig:3d-rashba-lengthscales} summarizes the transverse
pair-size and coherence-length components across the
$
1/(k_\mathrm{F}a_s)$--$m\alpha/k_\mathrm{F}
$
parameter space. Since the transverse and longitudinal components of
each length scale coincide in the zero-SOC limit, we drop the $\perp$
subscript when discussing zero-SOC results below.
On the BCS side, where
$
\mu\simeq\varepsilon_\mathrm{F}
$
and
$
\Delta_0/\varepsilon_\mathrm{F}\rightarrow0,
$
the two length scales behave as
$
k_\mathrm{F}^2\xi_{\mathrm{Cp}}^2 \simeq\varepsilon_\mathrm{F}^2/(6\Delta_0^2)
$
and
$
k_\mathrm{F}^2\xi_{0}^2 \simeq\varepsilon_\mathrm{F}^2/(9\Delta_0^2),
$
and both therefore diverge as the pairing gap vanishes.
In the zero-SOC BEC limit, they instead approach
$
k_\mathrm{F}^2\xi_{0}^2\rightarrow3\pi/(16k_\mathrm{F}a_s)
$
and
$
k_\mathrm{F}^2\xi_{\mathrm{Cp}}^2\rightarrow(k_\mathrm{F}a_s)^2/6,
$
consistent with the
literature~\cite{sademelo93, engelbrecht97, pistolesi94, pistolesi96}.
Note that the pair size differs from this literature value by a factor
of 3, which arises because we report a single Cartesian component
rather than the trace of the pair-size tensor. In particular, the
BEC-limit coherence length follows from the effective Gross-Pitaevskii
description of a dilute gas of composite bosons. For bosons with
density $n_\mathrm{B}$, mass $m_\mathrm{B}$, and interaction strength
$
U_\mathrm{BB} = 4\pi a_\mathrm{BB}/m_\mathrm{B},
$
the coherence length is given by
$
\xi_0^2=1/(4m_\mathrm{B} U_\mathrm{BB} n_\mathrm{B})
= 1/(16\pi n_\mathrm{B} a_\mathrm{BB}).
$
Within the mean-field approximation, $n_\mathrm{B}=n/2$,
$m_\mathrm{B}=2m$, and $a_\mathrm{BB}=2a_s$~\cite{sademelo93, engelbrecht97, pistolesi94, pistolesi96},
which directly gives the expression above. Thus, the opposing BEC-limit
behaviors of the two length scales reflect their different physical
content. As $1/(k_\mathrm{F}a_s)$ increases, stronger binding reduces
the internal size of the molecules, causing the pair size to vanish.
By contrast, the effective interaction between the composite bosons
becomes weaker, so collective variations of the condensate extend over
increasingly longer distances and the coherence length grows,
consistent with the behavior of a weakly-interacting BEC.

At finite SOC, the transverse pair size decreases monotonically over the
calculated parameter range, consistent with the progressive localization
of the pairs. Combining the strong-SOC two-body bound-state result
of Ref.~\cite{yu2011socfeshbach} with our pair size expression,
we obtain the universal strong-SOC asymptote
$
k_\mathrm{F}^2 \xi_{\mathrm{Cp}, \perp}^2
\simeq 0.460\left(\frac{k_\mathrm{F}}{m\alpha}\right)^2.
$
This asymptote is independent of $a_s$ as long as $U\neq0$, consistent with the
diverging-density-of-states argument of Sec.~\ref{sec:twobody}.
The convergence to this asymptote is fastest
near unitarity, whereas the BCS- and BEC-side results show larger
finite-SOC corrections, approaching it from above and below, respectively.
The overall finite-SOC behavior obtained numerically also agrees with Ref.~\cite{yu2014coherence}.
The transverse coherence length, by contrast, exhibits a qualitatively
different SOC dependence: on the weak- and intermediate-coupling sides,
it initially decreases, reaches a minimum, and then increases again at
stronger SOC. The position of this minimum closely follows the condition
$\mu\simeq-m\alpha^2/2$, where the chemical potential approaches the
minimum of the $-$-helicity Rashba band. As discussed in
Sec.~\ref{sec:single_particle}, the corresponding noninteracting FS
collapses onto the Rashba ring at this point.
Here, this surface is defined by $\xi_{s\mathbf k}=0$ at the
self-consistent chemical potential. The associated changes in the
low-energy helicity-band structure affect the gradient response entering
the coherence length, while the paired-state excitation spectrum remains
gapped. The stronger-SOC regime, in contrast, is increasingly governed by
bound-pair physics, where the Gaussian theory again reduces to an effective
Gross-Pitaevskii form for the weakly-interacting ``rashbon'' BEC.
Using the bound-state energy and the transverse effective mass
obtained in Ref.~\cite{yu2011socfeshbach} in the strong-SOC expansion
gives
$
k_\mathrm{F}^2 \xi_{0,\perp}^2 \simeq 0.3476\,\frac{m\alpha}{k_\mathrm{F}},
$
which is in very good agreement with our numerics.
The transverse coherence length squared therefore increases linearly
with SOC in this regime.

The competition between the intraband and interband contributions
also affects the SOC dependence on the BEC side. For example,
at $1/(k_\mathrm{F}a_s)=1$, the intraband component
$\xi_{0,\perp}^{2,\mathrm{intra}}$ initially decreases and reaches
a minimum near $m\alpha/k_\mathrm{F}\simeq0.48$, while the total
$\xi_{0,\perp}^2$ increases monotonically over the calculated SOC
range. The interband contribution grows sufficiently rapidly to
compensate for the initial intraband decrease, so that the minimum
is absent from the total. Thus, although the intraband contribution
remains dominant, the quantum-geometric term can change the qualitative
SOC dependence of the coherence length.

On the weak-coupling side, both interband contributions display a
distinct structure near the $\mu=0$ contour. As the pairing gap
decreases, these terms become increasingly sensitive to the
low-momentum states near the underlying helicity FSs.
After summing over helicity, the pair-size kernel in
Eq.~\eqref{eq:many_body_pair_size_inter} is proportional to
$(1/E_{+,\mathbf k}-1/E_{-,\mathbf k})^2$. The near-degeneracy of the
two quasiparticle energies at low momentum therefore suppresses the
interband pair-size contribution and produces a shallow minimum. The
coherence-length kernel behaves differently: for Rashba SOC,
$
(\xi_{s\mathbf k}-\xi_{-s,\mathbf k})^2 = 4 d_{\mathbf k}^2
\propto k_\perp^2,
$
which compensates the $1/k_\perp^2$ behavior of the transverse quantum
metric. This leaves a finite low-momentum contribution that becomes
more pronounced as the pairing gap decreases, producing the
enhancement near the same $\mu=0$ contour.

For the pair size,
$
\xi_{\mathrm{Cp},\perp}^{2,\mathrm{inter}}/\xi_{\mathrm{Cp},\perp}^2
$
reaches a maximum of approximately $0.322$ on the calculated grid, at
$1/(k_\mathrm{F}a_s)=-0.20$ and $m\alpha/k_\mathrm{F}\simeq0.967$. For
the coherence length,
$
\xi_{0,\perp}^{2,\mathrm{inter}}/\xi_{0,\perp}^2
$
reaches a maximum of approximately $0.237$ at
$1/(k_\mathrm{F}a_s)\simeq-1.23$ and $m\alpha/k_\mathrm{F}\simeq1.92$.
These ratios refer to the squared lengths, for which the intraband and
interband contributions are additive.

\begin{figure*}[htb]
\centering
\includegraphics[width=\textwidth]{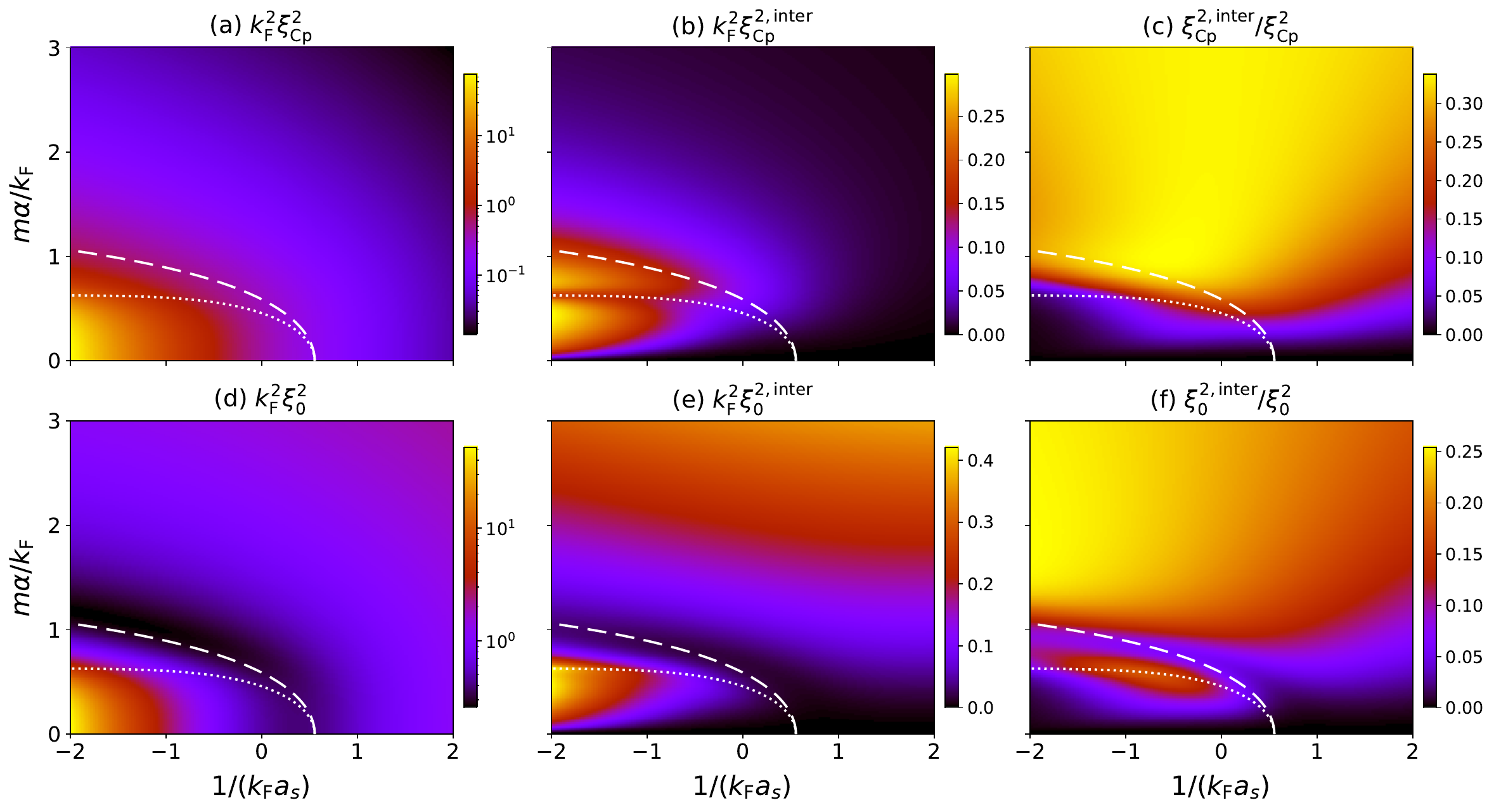}
\caption{
Pair-size and coherence-length components for the 3D Weyl model as
functions of $1/(k_\mathrm{F}a_s)$ and $m\alpha/k_\mathrm{F}$.
Panels (a)--(c) show the total pair size $k_\mathrm{F}^2\xi_{\mathrm{Cp}}^2$,
its interband contribution $k_\mathrm{F}^2\xi_{\mathrm{Cp}}^{2,\mathrm{inter}}$,
and the relative interband contribution
$\xi_{\mathrm{Cp}}^{2,\mathrm{inter}}/\xi_{\mathrm{Cp}}^2$, respectively.
Panels (d)--(f) show the corresponding quantities for the coherence length,
$k_\mathrm{F}^2\xi_0^2$, $k_\mathrm{F}^2\xi_0^{2,\mathrm{inter}}$, and
$\xi_0^{2,\mathrm{inter}}/\xi_0^2$. Logarithmic color scales are used for
the total quantities, and linear scales for the interband and relative
interband contributions. White dashed and dotted contours indicate
$\mu=-m\alpha^2/2$ and $\mu=0$, respectively.
}
\label{fig:3d-weyl-lengthscales}
\end{figure*}
\subsection{Three-dimensional system with Weyl SOC}
\label{sec:3d-weyl}

For the 3D Weyl model, rotational invariance implies
$
(\xi_{\mathrm{Cp}}^2)_{xx}=(\xi_{\mathrm{Cp}}^2)_{yy}
=(\xi_{\mathrm{Cp}}^2)_{zz}
$
and
$
(\xi_0^2)_{xx}=(\xi_0^2)_{yy}=(\xi_0^2)_{zz},
$
with all off-diagonal components of both tensors vanishing; we
accordingly define the isotropic scalar components $\xi_{\mathrm{Cp}}^2$
and $\xi_0^2$, together with their intraband and interband parts. Unlike
the 3D Rashba case, the helicity states here depend on all three
momentum components, so the quantum metric contributes equally to
every diagonal direction rather than only to the transverse plane.
We present the resulting dimensionless quantities
$k_\mathrm{F}^2\xi_{\mathrm{Cp}}^2$ and $k_\mathrm{F}^2\xi_0^2$
together with their quantum-geometric interband contributions and
relative fractions.

Figure~\ref{fig:3d-weyl-lengthscales} summarizes the pair-size and
coherence-length components across the
$
1/(k_\mathrm{F}a_s)$--$m\alpha/k_\mathrm{F}
$
parameter space. In the zero-SOC limit, both length scales recover
the weak- and strong-coupling limits given in Sec.~\ref{sec:3d-rashba}.
Finite Weyl SOC reduces the total pair-size component monotonically
over the entire calculated range, consistent with the progressive
localization of the pairs. In the strong-SOC limit, combining the Weyl
bound-state result of Ref.~\cite{he2012weyl} with our pair size
expression gives
$
k_\mathrm{F}^2\xi_{\mathrm{Cp}}^2
\simeq\frac{1}{4(m\alpha/k_\mathrm{F})^2}.
$
The coherence length again shows a qualitatively different SOC
dependence, initially decreasing, reaching a minimum, and increasing
again on the weak- and intermediate-coupling sides. As in the 3D
Rashba model, the minimum tracks the point at which the chemical
potential reaches the bottom of the lower-helicity band,
$\mu\simeq-m\alpha^2/2$, here corresponding to the sphere of minima
$k=m\alpha$ rather than the Rashba ring. In the strong-SOC limit, the
system enters the weakly-interacting ``rashbon'' BEC regime,
and using the effective Gross-Pitaevskii result of Sec.~\ref{sec:3d-rashba}
with $n_\mathrm{B}=n/2$ and the asymptotic bound-pair scattering length of
Ref.~\cite{he2012weyl} gives
$
k_\mathrm{F}^2\xi_0^2\simeq\frac{7\pi}{8(4+\sqrt{2})}
\frac{m\alpha}{k_\mathrm{F}}
\simeq0.5077\,\frac{m\alpha}{k_\mathrm{F}},
$
in very good agreement with our numerics; the coherence length
squared therefore again increases linearly with SOC in this regime.

Both interband contributions exhibit the same low-SOC structure near
the $\mu=0$ contour as in the 3D Rashba model, arising from the same
low-momentum mechanisms discussed in Sec.~\ref{sec:3d-rashba}, now
within the isotropic spherical helicity geometry of the Weyl model:
the pair-size feature weakens as the interaction strength increases,
while the coherence-length contribution is instead suppressed near the
bottom of the lower-helicity band. On the calculated grid,
$
\xi_{\mathrm{Cp}}^{2,\mathrm{inter}}/\xi_{\mathrm{Cp}}^2
$
reaches a maximum of approximately $0.338$ at $1/(k_\mathrm{F}a_s)=-0.55$
and $m\alpha/k_\mathrm{F}=1.00$, while
$
\xi_0^{2,\mathrm{inter}}/\xi_0^2
$
reaches a maximum of approximately $0.254$ at $1/(k_\mathrm{F}a_s)=-2.00$
and $m\alpha/k_\mathrm{F}\simeq1.767$.

\begin{figure*}[htb]
\includegraphics[width=\textwidth]{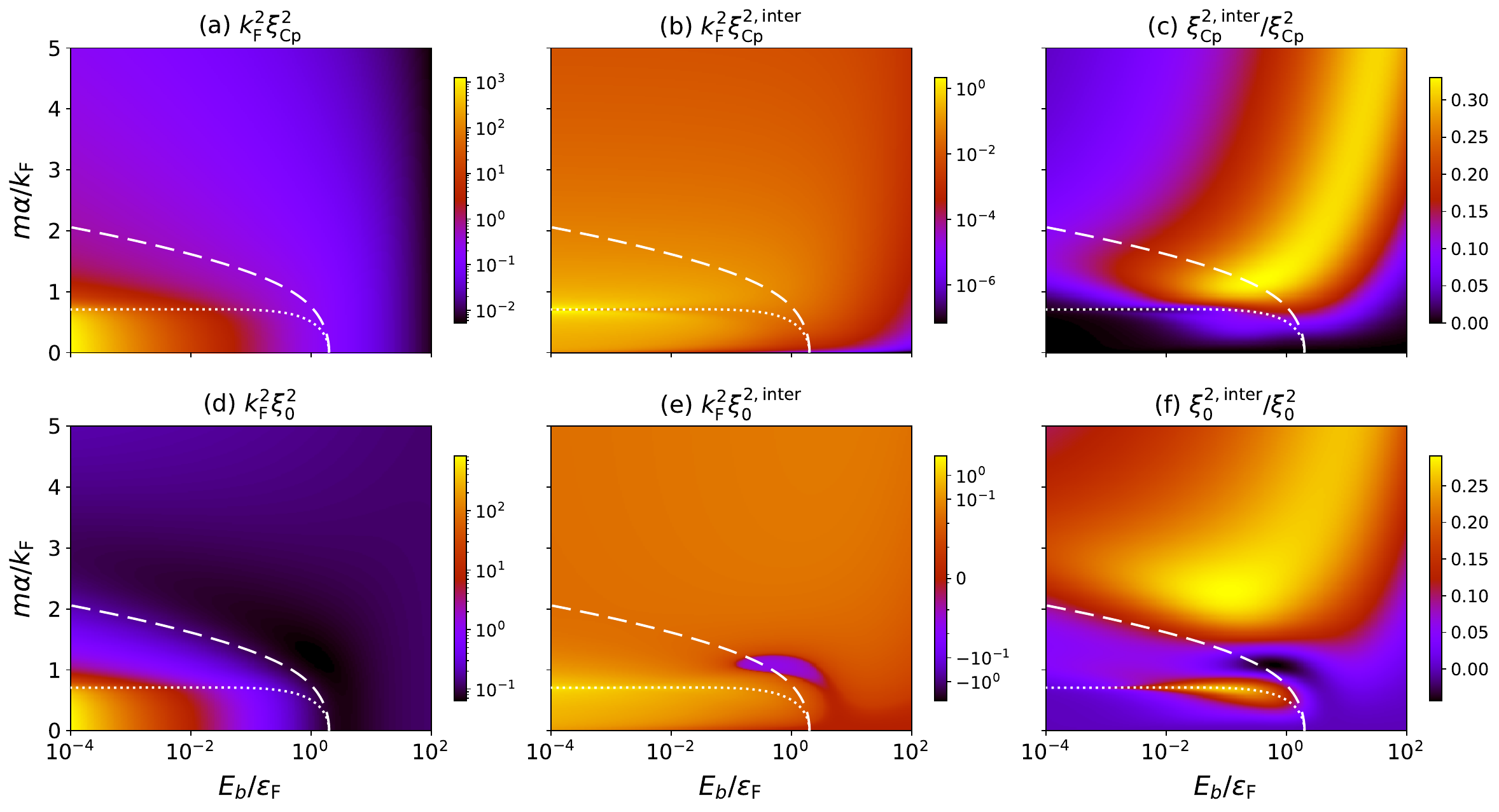}
\caption{
Pair-size and coherence-length components for the
2D Rashba model as functions of $E_b/\varepsilon_\mathrm{F}$ and
$m\alpha/k_\mathrm{F}$.
Panels (a)--(c) show the total pair size $k_\mathrm{F}^2\xi_{\mathrm{Cp}}^2$,
its interband contribution $k_\mathrm{F}^2\xi_{\mathrm{Cp}}^{2,\mathrm{inter}}$,
and the relative interband contribution
$\xi_{\mathrm{Cp}}^{2,\mathrm{inter}}/\xi_{\mathrm{Cp}}^2$, respectively.
Panels (d)--(f) show the corresponding quantities for the coherence length,
$k_\mathrm{F}^2\xi_0^2$, $k_\mathrm{F}^2\xi_{0}^{2,\mathrm{inter}}$,
and $\xi_{0}^{2,\mathrm{inter}}/\xi_0^2$. Logarithmic color scales are
used for the total quantities and $\xi_{\mathrm{Cp}}^{2,\mathrm{inter}}$,
a symmetric-logarithmic scale for $\xi_0^{2,\mathrm{inter}}$, and linear
scales for the relative interband contributions.
White dashed and dotted contours indicate $\mu=-m\alpha^2/2$ and $\mu=0$,
respectively.
}
\label{fig:2d-rashba-lengthscales}
\end{figure*}
\subsection{Two-dimensional system with Rashba SOC}
\label{sec:2d-rashba}

Having established the general trends in 3D, we now turn to the 2D
Rashba model, where the reduced dimensionality gives rise to
logarithmically divergent SOC-driven asymptotics not present in 3D.
Rotational symmetry in the plane implies
$(\xi_{\mathrm{Cp}}^2)_{xx}=(\xi_{\mathrm{Cp}}^2)_{yy}$ and
$(\xi_0^2)_{xx}=(\xi_0^2)_{yy}$, with all off-diagonal components
vanishing; we accordingly define the scalar components $\xi_{\mathrm{Cp}}^2$
and $\xi_0^2$, together with their intraband and interband parts, as
before.

Figure~\ref{fig:2d-rashba-lengthscales} summarizes the pair-size and
coherence-length components across the
$E_b/\varepsilon_\mathrm{F}$--$m\alpha/k_\mathrm{F}$ parameter space. In
the absence of SOC and in the weak-binding limit, where
$\mu\simeq\varepsilon_\mathrm{F}$ and
$\Delta_0^2\simeq2\varepsilon_\mathrm{F}E_b$, the two length scales behave
as $k_\mathrm{F}^2\xi_{\mathrm{Cp}}^2\simeq\varepsilon_\mathrm{F}/(8E_b)$
and $k_\mathrm{F}^2\xi_{0}^2\simeq\varepsilon_\mathrm{F}/(12E_b)$, both
diverging as $E_b\rightarrow0$. In the strong-binding limit,
they instead approach
$
k_\mathrm{F}^2\xi_{\mathrm{Cp}}^2\simeq2\varepsilon_\mathrm{F}/(3E_b)
$
and
$
k_\mathrm{F}^2\xi_{0}^2\simeq1/8.
$
As in 3D, the pair size continues to
shrink with the internal size of the tightly-bound molecules, while
the coherence length saturates at a finite value.
Within the effective Gross-Pitaevskii description to which the Gaussian
fluctuation theory reduces in this limit, this saturation reflects the
fact that composite bosons in 2D interact through a constant,
density-independent repulsion.

At finite SOC, the pair size decreases overall as either the binding
energy or the SOC strength is increased. In the strong-SOC limit,
combining the asymptotic two-body bound-state result of
Ref.~\cite{he2012rashba2d} with our pair size expression gives
$
k_\mathrm{F}^2\xi_{\mathrm{Cp}}^2\simeq\frac{1}{4(m\alpha/k_\mathrm{F})^2}
\big[1+\frac{[\ln(m\alpha^2/E_b)+2]^2}{\pi^2}\big].
$
As in the 3D models this asymptote holds for any $U\neq0$, though in
2D a bound state already exists at arbitrarily weak coupling without SOC,
so the ring of minima here sharpens rather than creates this tendency,
driving the system into the same weakly-interacting ``rashbon'' BEC
regime found in the 3D models.
The coherence length exhibits a different SOC dependence: at fixed
binding energy, it initially decreases, reaches a minimum at
intermediate SOC, and subsequently increases again at stronger SOC. In
the asymptotic limit $m\alpha^2/E_b\gg1$, substituting the large-SOC
forms of the bound-state energy and bound-pair effective mass from
Ref.~\cite{he2012rashba2d} gives
$
k_\mathrm{F}^2 \xi_0^2 \simeq \frac{1}{48}[ \ln(m\alpha^2/E_b)+2 ],
$
an asymptote our numerical results approach only gradually.

On the weak-binding side, both interband contributions display a sharp
feature near $m\alpha/k_\mathrm{F}\simeq1/\sqrt{2}$, where the
fixed-density condition gives $\mu=0$ in the limit $E_b\to0$. The
resulting suppression of the pair-size contribution and enhancement of
the coherence-length contribution follow from the same low-momentum
mechanisms discussed in Sec.~\ref{sec:3d-rashba}, here made
particularly sharp by the reduced dimensionality.
Away from the weak-binding region, the coherence-length interband
contribution is strongly suppressed at intermediate SOC and, at
intermediate binding energies, changes sign: it develops a shallow
negative minimum near $m\alpha/k_\mathrm{F}\simeq1.0$--$1.2$ before
becoming positive again at stronger SOC. This sign change originates
from the spectral kernel multiplying the quantum metric in
Eq.~\eqref{eq:coherence_gradient_inter}, which is not positive
definite, so contributions from different momentum regions can
partially cancel, allowing the interband term either to enhance or to
reduce the total coherence-length component relative to its intraband
part. This behavior is qualitatively distinct from the 3D Rashba and
Weyl models, where the interband coherence-length contribution remains
positive throughout the calculated parameter range: in 2D, the quantum
metric can therefore act to shorten the coherence length below its
intraband value, rather than merely adding to it.

On the calculated grid,
$
\xi_{\mathrm{Cp}}^{2,\mathrm{inter}}/\xi_{\mathrm{Cp}}^2
$
reaches a maximum of approximately $0.33$ at
$E_b/\varepsilon_\mathrm{F}\simeq0.383$ and $m\alpha/k_\mathrm{F}=1.2$.
For the coherence length,
$\xi_{0}^{2,\mathrm{inter}}/\xi_0^2$ reaches a minimum of approximately
$-0.044$ at $E_b/\varepsilon_\mathrm{F}\simeq0.639$ and
$m\alpha/k_\mathrm{F}\simeq1.083$. Here, the interband term is negative,
with a magnitude equal to approximately 4.4\% of the total squared
coherence length. The ratio reaches a maximum of
approximately $0.290$ at $E_b/\varepsilon_\mathrm{F}\simeq0.107$ and
$m\alpha/k_\mathrm{F}\simeq2.283$.

\subsection{Comparison across SOC geometries}
\label{sec:comparison}

On the weak-coupling side, the three models share the same sequence as
the SOC strength is increased at fixed attraction: both length scales
decrease as the low-energy states move toward the minima of the
$-$-helicity band, which form a sphere in the Weyl model and a ring in
both Rashba models; the underlying noninteracting FS collapses onto
this manifold at $\mu=-m\alpha^2/2$; and beyond the collapse the pair
size continues to decrease with the internal extent of the bound state,
whereas the coherence length grows as the interaction between
``rashbons'' weakens. The pair size is therefore monotonic, whereas the
coherence length has a minimum. In the 3D models the minimum lies close
to the collapse, and both occur at weaker SOC for Weyl than for Rashba
coupling: at unitarity, the collapse and the minimum are at
$m\alpha/k_\mathrm{F}\simeq0.59$ and $0.68$ for Weyl SOC, and at $0.86$
and $0.90$ for 3D Rashba SOC. In 2D the minimum lies well beyond the
collapse, e.g., at $1.85$ versus $1.31$ for
$E_b/\varepsilon_\mathrm{F}=0.1$.

These differences follow from the geometry of the manifold of minima.
For a dispersion that is quadratic in the $c$ directions normal to the
manifold, where $c$ is its codimension, the density of states behaves
as $(E-E_{\min})^{c/2-1}$, with $E_{\min}=-m\alpha^2/2$ the bottom of
the $-$-helicity band. It diverges as $(E-E_{\min})^{-1/2}$ for the
Weyl sphere and the 2D Rashba ring ($c=1$), whereas the additional
$k_z$ dispersion of the 3D Rashba ring ($c=2$) gives a finite density
of states at the band bottom. The Weyl model, with the larger low-energy
density of states, reaches the band bottom at weaker SOC. Beyond the
collapse, the strong-SOC asymptotes of
Secs.~\ref{sec:3d-rashba}--\ref{sec:2d-rashba} show that
$k_\mathrm{F}^2\xi_0^2$ grows linearly with $m\alpha/k_\mathrm{F}$ in
both 3D models, with prefactors $0.508$ for Weyl SOC and $0.348$ for
the transverse 3D Rashba component set by the ``rashbon'' mass and
scattering length, whereas in 2D it grows only logarithmically; this
slower growth is why the 2D minimum lies well beyond the collapse.

The momentum-integrated geometric contributions to both squared length
tensors are isotropic for Weyl SOC, whereas for 3D Rashba SOC they
enter only the transverse components, since the helicity spinors depend
on the direction of $\mathbf k$ for Weyl SOC and only on the azimuthal
angle of the transverse momentum for Rashba SOC. The peak geometric
fractions of the squared coherence length are nevertheless close, about
25\% for Weyl SOC and 24\% for the transverse component in 3D Rashba
SOC. The two Rashba models share the same in-plane quantum metric, yet
the interband contribution to the coherence length remains nonnegative
throughout the calculated range in 3D while it is negative over part of
the 2D crossover: the metric enters
Eq.~\eqref{eq:coherence_gradient_inter} multiplied by a spectral kernel
that is not positive definite, and this kernel and its integration over
the additional $k_z$ direction differ between the two models.

\section{Comparison with the literature}
\label{sec:literature}

The literature on spin-orbit-coupled Fermi gases contains different
definitions of pairing-related length scales, which do not correspond
directly to the pair size and coherence length considered here.
Refs.~\cite{yu2011socfeshbach,yu2014coherence} define a Cooper-pair size
from the ratio $\mathcal{V}_{s\mathbf{k}}/\mathcal{U}_{s\mathbf{k}}$
obtained by extracting a two-particle component from the coherent BCS
ground state. We note that this construction is not defined through a
physical many-body correlation function. This distinction is particularly
apparent in the BCS regime, where the resulting length remains of
the order of the interparticle spacing $1/k_\mathrm{F}$,
whereas the physical pair size diverges with
$
k_\mathrm{F} \xi_{\mathrm{Cp}} \propto \varepsilon_\mathrm{F}/\Delta_0,
$
as $\Delta_0\rightarrow0$. This unphysicality is also noted and emphasized
in Ref.~\cite{yu2014coherence}.
We instead define the average Cooper-pair size through the localization
tensor of the spin-resolved anomalous amplitude. At the mean-field
level, our $\xi_{\mathrm{Cp}}$ is identical to the quantity called the
coherence length in Ref.~\cite{yu2014coherence}, and the corresponding
numerical results are in agreement.
We separately define the coherence length $\xi_0$ from the static,
long-wavelength behavior of amplitude fluctuations of the order
parameter; this quantity has no direct counterpart in these earlier
studies. We further decompose both $\xi_{\mathrm{Cp}}$ and $\xi_0$ into
intraband and quantum-geometric interband contributions and identify the
latter with the quantum metric of the helicity states. This reveals and
quantifies the quantum-geometric contribution to each length scale
throughout the interaction-SOC parameter space.

A similar terminological ambiguity appears in the multiband Hubbard-model
literature, where quantum-metric controlled length scales have been
extracted from different correlation functions and given different names.
Ref.~\cite{hu23} extracts a length from the pair-correlation function of
a flat-band model projected onto the flat band, identifies it with the
superconducting coherence length, and interprets its quantum-metric lower
bound as the minimal size of a Cooper pair. Ref.~\cite{thumin24} instead
extracts characteristic lengths from the real-space decay of the normal
and anomalous correlation functions, using the terms coherence length and
Cooper-pair size interchangeably, and finds no simple relation between
these lengths and the quantum metric. The localization-tensor
formulation of Refs.~\cite{iskin25, elden26}, on the other hand,
defines the pair size through the internal extent of the pair wave
function, and Ref.~\cite{elden26} compares it directly with the
Gaussian-fluctuation coherence length within the same model. The two
length scales behave qualitatively differently in the dilute flat-band
limit: the pair size remains finite and is controlled by the quantum
metric, whereas the coherence length diverges as the particle density
vanishes~\cite{iskin24c, iskin25, elden26}. We also note that the
zero-temperature coherence length obtained from the $\mathbf q\to\mathbf 0$
expansion of the amplitude channel is well defined only where the
gradient-coefficient tensor $C_{ij}$ is positive definite, i.e.,
where all principal components of $(\xi_0^2)_{ij}$ are positive.
In lattice models this condition can fail at intermediate fillings
and couplings, where $\xi_0^2$ turns negative~\cite{iskin24c, elden26}.
This failure is not merely formal: a negative $C_{ij}$ signals that the
static amplitude channel is minimized at a finite wave vector rather
than at $\mathbf q=\mathbf 0$, so that the coherence length must instead
be extracted from an expansion around that finite-$\mathbf q$ minimum,
with density fluctuations treated on the same footing as the pairing
fluctuations~\cite{elden26}. The condition holds in the dilute limit,
where the lattice problem reduces to the continuum crossover considered
here, and it is satisfied throughout the parameter space of the present
model. Together with our results for spin-orbit-coupled Fermi gases,
these findings reinforce the distinction between the internal extent of
Cooper pairs and the collective spatial response of the order parameter.

\section{Conclusion and Outlook}
\label{sec:conc}

We showed that, in the spin-orbit-coupled BCS-BEC crossover, the pair
size and the coherence length diverge markedly, and in qualitatively
different ways, once the system is taken away from the BCS regime, where
the two quantities almost coincide. Both admit the same decomposition
into an intraband contribution, set by the helicity-band dispersion, and
a quantum-geometric interband contribution, governed entirely by the
quantum metric, which we established for the exact two-body bound state
and extended to the mean-field Cooper-pair size and the
Gaussian-fluctuation coherence length at zero temperature. Across all
three SOC geometries, the pair size decreases monotonically with
increasing SOC, whereas the coherence length develops a pronounced
minimum, associated with the collapse of the noninteracting FS onto the
sphere or ring of helicity-band minima, before growing again as the composite
bosons become weakly interacting; in all three models this growth is
consistent with the effective Gross-Pitaevskii description of a dilute
``rashbon'' BEC, though the logarithmic 2D asymptote is approached only
gradually. The quantum-geometric contribution is subdominant but
nonnegligible, reaching up to about 30\% of both the squared pair
size and the squared coherence length, and in the 2D case it even
changes sign. The differences among the three geometries were traced to
the codimension of the manifold of helicity-band minima, the direction
dependence of the helicity spinors, and the dimensionality of the
``rashbon'' BEC (Sec.~\ref{sec:comparison}).
We emphasize that the two length scales are physically distinct objects,
a spatial second moment of the pair wavefunction versus the gradient
stiffness of the order parameter, and that the quantity called the
coherence length in earlier work on spin-orbit-coupled Fermi
gases~\cite{yu2014coherence} corresponds to the former rather than the
latter (Sec.~\ref{sec:literature}).

These results made concrete, in a fully dispersive and numerically
tractable setting, a distinction that had remained difficult to pin
down in the flat-band superconductivity
literature~\cite{hu23, thumin24, iskin24c, iskin25, elden26}: the two
length scales need not track one another, even though both originate from
the same pairing correlations and both receive a genuine geometric
contribution. Here the separation was transparent because a finite
intraband dispersion coexisted with the quantum-geometric term, allowing
the two contributions to be cleanly disentangled. This separation is
physically meaningful in the same sense as that of the superfluid
weight into conventional and geometric parts: although only the total
is observable, the decomposition identified which features of the
SOC dependence originated from the helicity dispersion and which from
the quantum metric of the helicity spinors, and here the latter left
qualitative signatures in the total. We expect this lesson,
that the coherence length is sensitive to the stiffness of the order
parameter and can therefore behave nonmonotonically even where the
pair size varies smoothly, to offer a useful reference point for
revisiting the flat-band problem, where the vanishing kinetic term
removes this convenient separation and has arguably contributed to the
conflicting conclusions reached in that literature.

On the theoretical side, extending this framework to lattice
realizations in which the helicity bands can be tuned continuously into
the exactly flat limit would allow the intraband/quantum-geometric
separation established here to be traced directly into the regime where
the flat-band controversy is sharpest. Given the ongoing experimental
progress in realizing both
SOC~\cite{galitski13, zhai15, zhang19, cheuk12, meng16, huang16}
and flat-band-like lattice
geometries~\cite{taie15, leung20, meng23, lebrat26}
with ultracold atoms, a direct measurement of these two quantities,
for instance through pair-correlation or collective-mode spectroscopy,
across the crossover studied here would provide a valuable and,
in principle, accessible test of the quantum-geometric contributions
identified in this work.

\begin{acknowledgments}
We acknowledge support from the U.S. Air Force Office of Scientific
Research (AFOSR) under Grant No.~FA8655-24-1-7391.
\end{acknowledgments}

\appendix

\section{Derivation of Eq.~\eqref{eq:coherence_gradient_inter}}
\label{app:interband-gradient}

At $\nu_\ell=0$, Eqs.~\eqref{eq:fluctuation_M11} and
\eqref{eq:fluctuation_M12} give
$
M_{11}(\mathbf q,0)+M_{12}(\mathbf q,0)
=\frac{V}{U}-\frac18\sum_{ss'\mathbf k}
( 1 - ss'\hat{\mathbf d}^{+}\cdot\hat{\mathbf d}^{-} )
\frac{E_s^+E_{s'}^-+\xi_s^+\xi_{s'}^--\Delta_0^2}
{E_s^+E_{s'}^-(E_s^++E_{s'}^-)},
$
where the overlap factor equals four times the singlet weight
$|\mathcal{A}^{\mathbf q}_{ss'\mathbf k}|^2$ of Sec.~\ref{sec:twobody}. The
$q^2$ expansions of this factor and of the energy-dependent fraction
determine the geometric and intraband contributions to $C_{ij}$,
respectively. Since $\hat{\mathbf d}_{-\mathbf k}=-\hat{\mathbf d}_{\mathbf k}$
for the SOC models considered here,
$\hat{\mathbf d}^{+}\cdot\hat{\mathbf d}^{-}
=-\hat{\mathbf d}_{\mathbf k+\mathbf q/2}\cdot\hat{\mathbf d}_{\mathbf k-\mathbf q/2}$.
Expanding
$
\hat{\mathbf d}_{\mathbf k\pm\mathbf q/2}
=\hat{\mathbf d}_{\mathbf k}
\pm\frac12\sum_i q_i\partial_i\hat{\mathbf d}_{\mathbf k}
+\frac18\sum_{ij}q_iq_j\partial_i\partial_j\hat{\mathbf d}_{\mathbf k}
+\mathcal O(\mathbf{q}^3)
$
and using the normalization identity
$\hat{\mathbf d}_{\mathbf k}\cdot\partial_i\partial_j\hat{\mathbf d}_{\mathbf k}
=-\partial_i\hat{\mathbf d}_{\mathbf k}\cdot\partial_j\hat{\mathbf d}_{\mathbf k}
=-2g^{\mathbf k}_{ij}$, we obtain
\begin{equation}
1-ss'\hat{\mathbf d}^{+}\cdot\hat{\mathbf d}^{-}
=1+ss'-ss'\sum_{ij}g^{\mathbf k}_{ij}q_iq_j+\mathcal O(\mathbf{q}^4).
\label{eq:app-overlap}
\end{equation}
The overlap is even in $\mathbf q$, and to quadratic order the quantum
metric transfers singlet weight from same-helicity to opposite-helicity
pairs.

The zeroth-order weight $1+ss'$, nonzero only for $s'=s$, multiplied by
the $q^2$ expansion of the energy-dependent factor gives, after
integrating second-derivative terms by parts over $\mathbf k$, the
intraband coefficient in Eq.~\eqref{eq:coherence_gradient_intra}. The
metric term multiplied by the energy-dependent factor at $\mathbf q=\mathbf 0$,
where $E_{s'}^-\to E_{s'\mathbf k}$ by evenness of the helicity
dispersions, gives the geometric part,
$
C_{ij}^{\mathrm{inter}}
=\frac18\sum_{ss'\mathbf k}ss'
\frac{E_{s\mathbf k}E_{s'\mathbf k}+\xi_{s\mathbf k}\xi_{s'\mathbf k}-\Delta_0^2}
{E_{s\mathbf k}E_{s'\mathbf k}(E_{s\mathbf k}+E_{s'\mathbf k})}\,g^{\mathbf k}_{ij}.
$
For $s'=s$ the fraction reduces to $\xi_{s\mathbf k}^2/E_{s\mathbf k}^3$,
so that separating the two helicity sectors gives
\begin{equation}
C_{ij}^{\mathrm{inter}}
=\frac18\sum_{s\mathbf k}
\Big[
\frac{\xi_{s\mathbf k}^2}{E_{s\mathbf k}^3}
-\frac{E_{s\mathbf k}E_{-s,\mathbf k}
+\xi_{s\mathbf k}\xi_{-s,\mathbf k}-\Delta_0^2}
{E_{s\mathbf k}E_{-s,\mathbf k}
(E_{s\mathbf k}+E_{-s,\mathbf k})}
\Big]g^{\mathbf k}_{ij},
\label{eq:app-separated}
\end{equation}
where both terms in the bracket are nonnegative. Using
$E_{s\mathbf k}^2-\xi_{s\mathbf k}^2=\Delta_0^2$ to write
$\xi_{s\mathbf k}^2/E_{s\mathbf k}^3=1/E_{s\mathbf k}-\Delta_0^2/E_{s\mathbf k}^3$
and
$
E_{s\mathbf k}E_{-s,\mathbf k}+\xi_{s\mathbf k}\xi_{-s,\mathbf k}-\Delta_0^2
=[(E_{s\mathbf k}+E_{-s,\mathbf k})^2
-(\xi_{s\mathbf k}-\xi_{-s,\mathbf k})^2-4\Delta_0^2]/2,
$
Eq.~\eqref{eq:app-separated} becomes
$
C_{ij}^{\mathrm{inter}}
=\sum_{s\mathbf k}
\big[
\frac1{16}\big(\frac1{E_{s\mathbf k}}-\frac1{E_{-s,\mathbf k}}\big)
-\frac{\Delta_0^2}{8E_{s\mathbf k}^3}
+\frac{(\xi_{s\mathbf k}-\xi_{-s,\mathbf k})^2+4\Delta_0^2}
{16E_{s\mathbf k}E_{-s,\mathbf k}(E_{s\mathbf k}+E_{-s,\mathbf k})}
\big]g^{\mathbf k}_{ij},
$
whose first term vanishes upon summation over $s$. The remaining terms
are Eq.~\eqref{eq:coherence_gradient_inter}.
This derivation also shows why the geometric contribution has no definite
sign: the same-helicity and opposite-helicity weight changes in
Eq.~\eqref{eq:app-overlap} enter with different energy-dependent factors,
so Eq.~\eqref{eq:app-separated} is a difference of two nonnegative terms
multiplying the positive-semidefinite metric. The geometric contribution
can therefore reduce the gradient coefficient $C_{ij}$, as found for the
diagonal components in part of the 2D Rashba parameter regime.

\bibliography{refs}

\end{document}